\documentclass[
reprint,
amsmath,
amssymb,
aps,
prb,
longbibliography,
noeprint
]{revtex4-2}
\usepackage{graphicx}
\usepackage{dcolumn}
\usepackage{bm}
\usepackage{comment}
\usepackage{needspace}
\usepackage{xcolor}
\usepackage[normalem]{ulem}
\usepackage{hyperref}
\hypersetup{
    colorlinks,
    linkcolor={red!50!black},
    citecolor={blue!50!black},
    urlcolor={blue!80!black}
}
\usepackage[capitalise]{cleveref}
\usepackage{hyperref} 

\newenvironment{myfont}{\fontfamily{phv}\selectfont}{\par}
\makeatletter
\renewcommand\frontmatter@abstractwidth{\dimexpr\textwidth-0.5in\relax}
\makeatother



\begin{document}

\preprint{APS/123-QED}

\title{\myfont Radio-frequency assisted switching in perpendicular magnetic tunnel junctions}

\author{\myfont Mark Hayward$^{1,}$}
\author{\myfont Salvatore Perna$^{2,}$}
\author{\myfont Massimiliano d'Aquino$^{2,}$}
\author{\myfont Claudio Serpico$^{2,}$}
\author{\myfont Wonjoon Jung$^{3,}$}
\author{\myfont Chunhui Dai$^{3\dagger}$}
\author{\myfont Patrick M. Braganca$^{3}$}
\author{\myfont Ilya N. Krivorotov$^{1,}$}
\email{Corresponding author email: ikrivoro@uci.edu\\$\dagger$ Current address: Samsung Semiconductor Inc, 3655 N. First Street, San Jose, CA 95134, USA}

\affiliation{\myfont $^1$ Department of Physics and Astronomy, University of California, Irvine, California, 92697, USA}
\affiliation{\myfont $^2$ Department of Electrical Engineering and Information Technology, University of Naples Federico II, Naples, Italy.}
\affiliation{\myfont $^3$ Western Digital Technologies, San Jose, California, 95119, USA}

\begin{myfont}

\begin{abstract}
\noindent\textbf{Abstract:} Spin-transfer torque magnetic random-access memory (STT-MRAM) relies on nanoscale magnetic tunnel junctions (MTJs) as its fundamental building blocks.
Next-generation STT-MRAM requires strategies that simultaneously improve switching energy efficiency and device endurance. 
Here, we present the first study of perpendicular STT-MRAM writing assisted by radio-frequency (RF) spin torque. 
We show that applying a small-amplitude RF pulse prior to a direct-current (DC) writing pulse enhances the MTJ switching probability, with the efficiency gain increasing at lower RF frequencies. 
This RF+DC writing scheme enables shorter DC pulses, thereby improving device endurance. 
Analytical and numerical modeling qualitatively reproduces the experimental trends, while quantitative discrepancies indicate that realistic MTJ properties beyond idealized models play an important role in RF-assisted switching.

\end{abstract}

\maketitle

\end{myfont}
\bigskip

\begin{large}
\noindent \textbf{\myfont Introduction}
\end{large}

STT-MRAM is a leading candidate for next-generation nonvolatile memory, combining sub-nanosecond switching \cite{rowlands2011, marins_de_castro2012, cubukcu2018, yamamoto2022}, CMOS compatibility \cite{rizzo2013, apalkov2016, worledge2024spin, khanal2021}, scalability to future technology nodes \cite{watanabe2018}, and radiation hardness \cite{montoya2020}. 
Significant progress has been made toward commercialization, yet the technology is currently limited to niche applications \cite{sakhare2018}. 
Advancing the technology to mainstream implementation requires overcoming the key challenges of relatively low energy efficiency, high write current density, and limited endurance \cite{worledge2024spin}. 
In perpendicular MTJs (p-MTJs), which constitute the core of contemporary STT-MRAM, the write current density must be kept high enough to guarantee reliable switching, but this comes at the cost of increased power consumption and the risk of tunnel barrier breakdown \cite{zhao2016}. 
Scaling the MTJ size below 20\,nm further exacerbates this trade-off by making the thermal stability and write reliability harder to balance \cite{perrissin2018}. 

The relatively high voltage pulses applied during the write operation accelerate degradation of the ultrathin MgO barrier, leading to reduced device lifetimes. 
This reliability challenge stresses the need for writing schemes that deliver switching at reduced voltage levels, without compromising retention or write error rates. 
An alternative, spin-orbit torque MRAM (SOT-MRAM), addresses the problem of dielectric breakdown by separating the read and write current paths \cite{miron2011, liu2012, yang2024, lopez-dominguez2023}. 
Although this decoupling enables fast switching without stressing the MgO barrier, the three-terminal geometry of SOT-MRAM imposes a large footprint that constrains storage density and integration. 
Therefore, innovation in the engineering of the STT-MRAM stack, device design, and write protocols is crucial for the development of energy-efficient STT-MRAM.

A straightforward strategy to reduce the write voltage in STT-MRAM is to increase the duration of the DC write pulse. 
However, this approach relying on thermally-assisted switching comes at the cost of higher write energy and reduced energy efficiency \cite{baek2019, sun2021}. 
The tradeoff between the write voltage amplitude and energy efficiency motivates the search for alternative write schemes \cite{kwiatkowski2021} that achieve reliable switching without compromising endurance or efficiency.\\
\indent An attractive approach is to combine the DC write pulse with an RF current pulse \cite{finocchio2006, cheng2013} -- a method inspired by studies of microwave-assisted magnetization reversal in magnetic media \cite{thirion2003, woltersdorf2007}. 
This RF-assisted strategy has previously been studied in in-plane MTJs and in all-metallic spin valves, where an enhanced switching probability was observed when the RF drive frequency was tuned to the ferromagnetic resonance (FMR) of the free layer (FL) \cite{cui2008resonant, florez2008}. 
In thermally unstable, superparamagnetic in-plane MTJs, a strong enhancement of random telegraph switching has been demonstrated at sub-GHz frequencies, where RF excitation drives low-dimensional chaotic dynamics of the FL \cite{montoya2019magnetization}.\\
\begin{figure*}[tbph]
\centering
 \includegraphics[width= 1.0\textwidth]{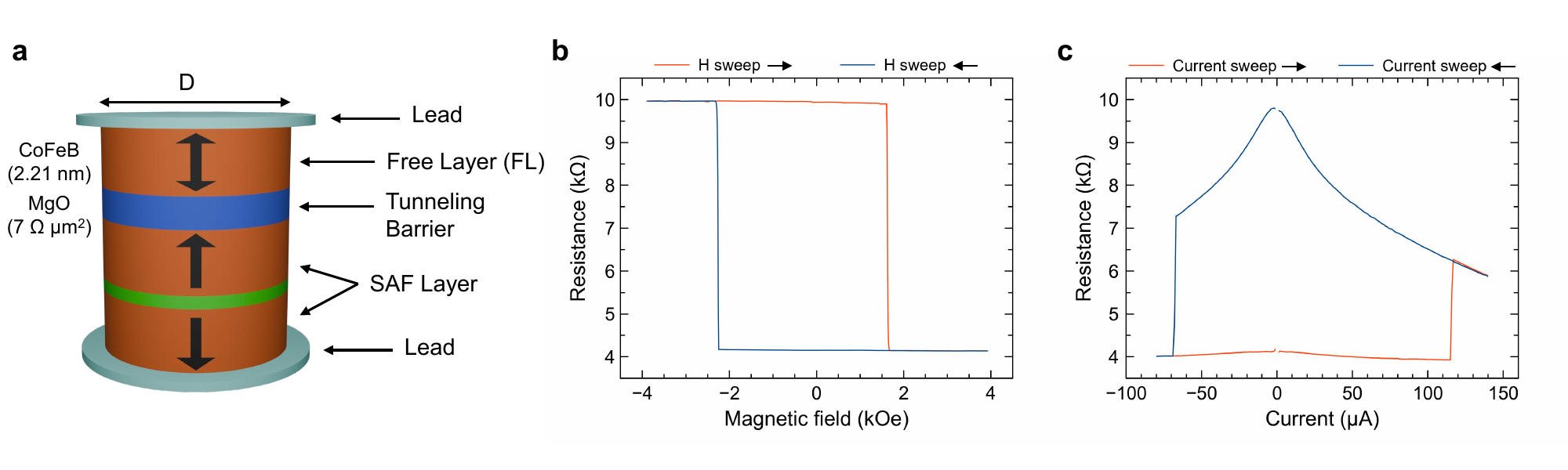}%
 \caption{{\textbf{Device schematics and switching characteristics of a p-MTJ. }\textbf{a} Schematic of a circular, nanoscale p-MTJ of diameter $D$ consisting of an SAF reference layer and a 2.21\,nm thick CoFeB FL separated by an MgO tunnel barrier with a resistance-area product of $7\, \Omega \, \mu$m$^2$.   
 Magnetic field \textbf{b} and current \textbf{c} hysteresis loops for a $D = 45$\,nm p-MTJ.} 
\label{fig:device_diagram_RvsH_RvsI}}%
 \end{figure*}
 \begin{figure*}[tbph]
\centering
 \includegraphics[width= 1.0\textwidth]{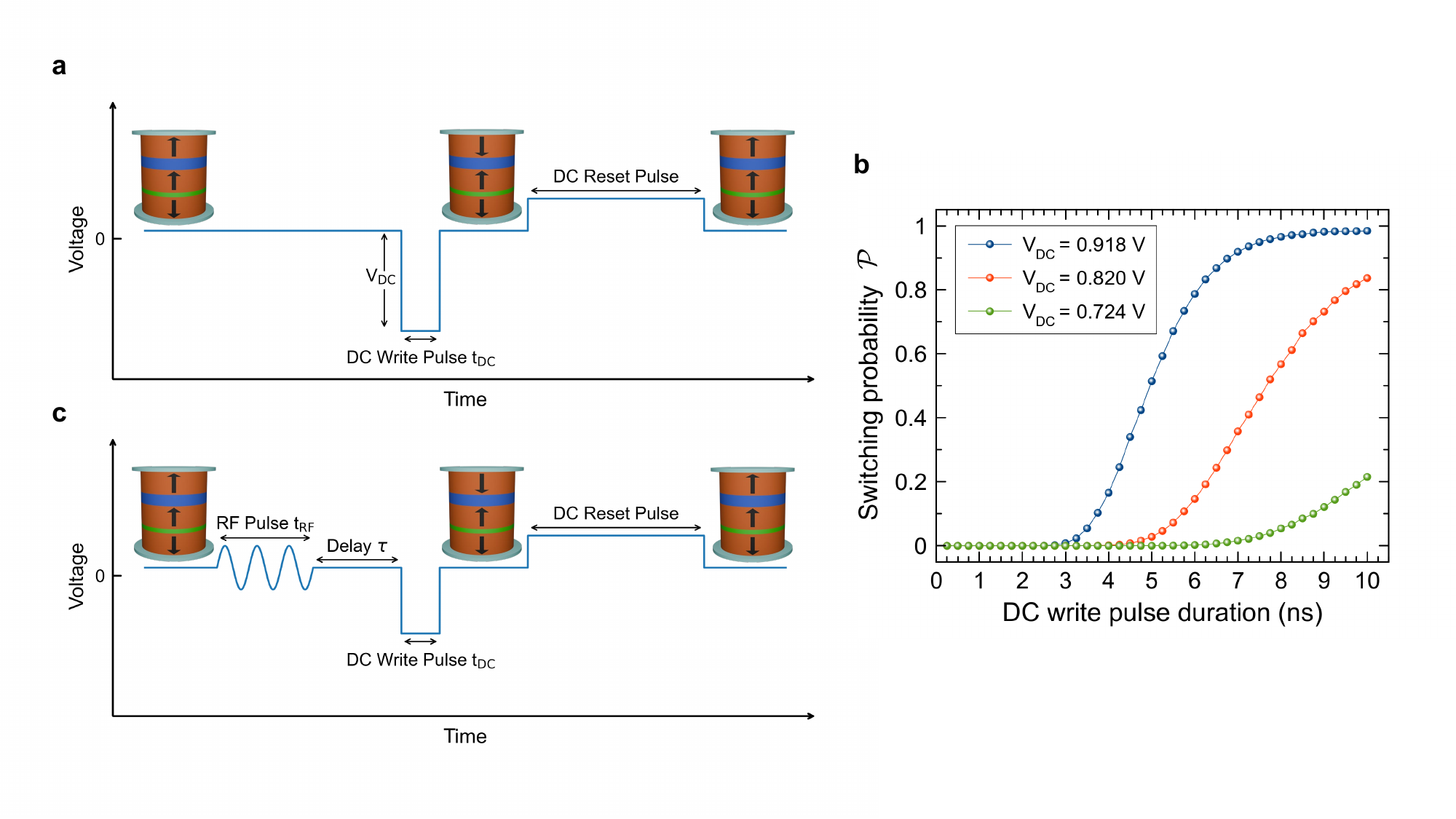}%
 \caption{{\textbf{Waveforms for current-induced p-MTJ switching.} \textbf{a} The waveform used for measurements of p-MTJ switching probability $\mathcal{P}$ in the conventional DC write pulse switching regime.  
 \textbf{b} $\mathcal{P}$ as a function of DC pulse duration $t_\mathrm{DC}$ for three pulse amplitudes measured using the pulse sequence in \textbf{a} for a $D = 45$\,nm device.
 \textbf{c} The waveform used for measurements of RF-assisted switching. 
  } 
\label{fig:pulse_chain_time_sweep}}%
 \end{figure*}
\indent Despite these promising results, RF-assisted switching \cite{carpentieri2010, taniguchi2016} in p-MTJs forming the basis of modern STT-MRAM technology has only been studied for RF frequencies near FMR \cite{suto2015}. 
In this work, we report experimental and theoretical studies of RF-assisted switching in thermally stable p-MTJs over a broad RF frequency range. 
We show that a small-amplitude RF pulse applied just before the DC write pulse can significantly increase the switching probability. 
This RF+DC writing protocol enables a reduction of the DC pulse duration, thereby improving device endurance.\\
\indent Furthermore, we find that the efficiency of RF assistance increases as the RF frequency is reduced. This is a notable technological advantage, since implementing sub-GHz RF drive circuitry in CMOS is simpler and more cost-effective than realizing GHz-frequency sources required for the FMR-frequency operation. We present measurements of RF-assisted switching across multiple p-MTJ diameters and compare these experimental results to an analytical model and numerical simulations.

\begin{large}
\noindent \textbf{\myfont Results}
\end{large}\\
\noindent\textbf{\myfont Device characterization}
\begin{figure*}[htbp]
\centering
 \includegraphics[width= 1.0\textwidth]{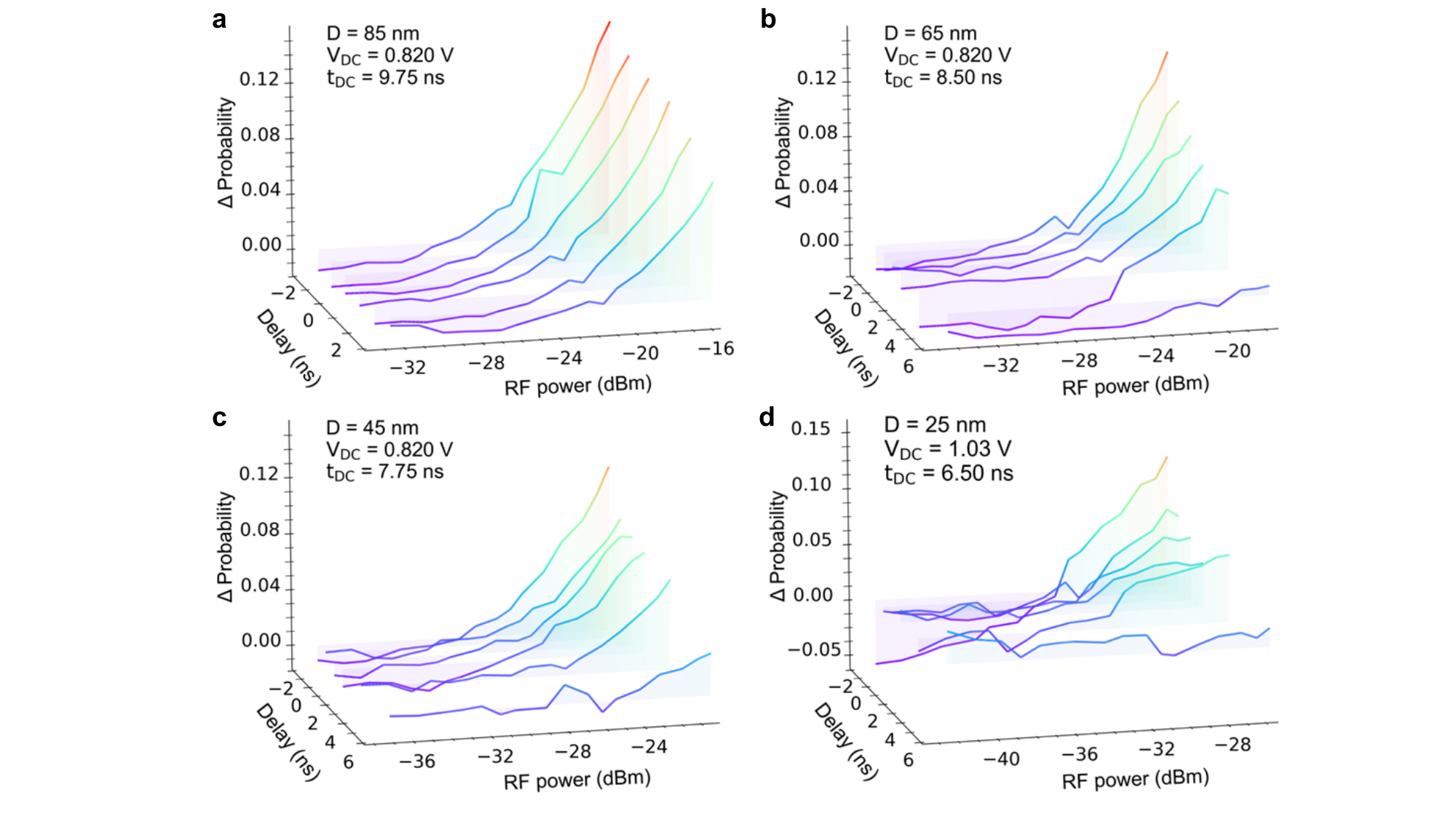}%
 \caption{{\textbf{RF-assisted increase in switching probability $\Delta \mathcal{P}$ versus RF power $P_\mathrm{RF}$ and RF-DC delay $\tau$.} 
 In these measurements, the RF pulse frequency $f_\mathrm{RF} = 0.1$\,GHz, and duration $t_\mathrm{RF}=30$\,ns are fixed. 
 The DC write pulse parameters for each p-MTJ are adjusted to achieve $\mathcal{P}=0.5$ in the absence of the RF pulse. 
 Four p-MTJ diameters are studied: \textbf{a} $D=85$\,nm, \textbf{b} $D=65$\,nm, \textbf{c} $D=45$\,nm, and \textbf{d} $D=25$\,nm.
 } 
\label{fig:Switching_multiple_devices}}%
 \end{figure*}
We investigate RF-assisted switching in circular nanoscale p-MTJs, schematically shown in \cref{fig:device_diagram_RvsH_RvsI}a.
The devices studied have diameters $D=85$, 65, 45, and 25\,nm, a CoFeB FL  thickness $d=2.21$\,nm, and an MgO tunneling barrier with resistance–area product of $7 \, \Omega \, \mu$m$^2$. The pinned layer is a synthetic antiferromagnet (SAF) \cite{mckinnon2022} (Methods). 
\Cref{fig:device_diagram_RvsH_RvsI}b,\,c show measurements of resistance as a function of out-of-plane magnetic field and quasi-static current for a $D = 45$\,nm p-MTJ. 
\indent Tunneling magnetoresistance (TMR) measurements, averaged over 135 devices for each diameter, show a weak dependence on device size, with a mean $\mathrm{TMR} = 130\,\%$.
Using material parameters extracted from spin-torque FMR measurements, the macrospin approximation for the dimensionless thermal stability at room temperature ($\Delta$) predicts $\Delta = 85$ for the $D=25$\,nm p-MTJs.
This diameter is close to the macrospin limit, where experimental values are expected to be comparable to the model predictions \cite{sun_spin-torque_2013}.
These measurements demonstrate the high thermal stability and large tunneling magnetoresistance of the devices.\\

\noindent\textbf{\myfont RF-assisted switching measurements}
\Cref{fig:pulse_chain_time_sweep}a illustrates the pulse sequence used for measurements of p-MTJ switching probability $\mathcal{P}$ ($0 \le \mathcal{P} \le 1$) in the conventional DC write pulse switching regime. 
In this paper, we study switching from the low-resistance parallel (P) to the high-resistance antiparallel (AP) state at room temperature in zero external magnetic field. 
A large-amplitude square DC write pulse is applied to the device in the P state, followed by the resultant resistance readout and a longer reset pulse returning the MTJ back to the initial P state.
The measurement is repeated $10^5$ times to reliably measure $\mathcal{P}$ for a given write pulse duration $t_\mathrm{DC}$ and amplitude $V_\mathrm{DC}$. 
\Cref{fig:pulse_chain_time_sweep}b shows the results of our measurements for a $D=45$\,nm device as a function of $t_\mathrm{DC}$ for three values of $V_\mathrm{DC}$. All voltage values reported in this paper are voltages applied across the MTJ (Supplementary Note 1).

For studies of RF-assisted switching, the DC write pulse is preceded by an RF pulse of duration $t_\mathrm{RF}$, frequency $f_\mathrm{RF}$ and root mean square (RMS) amplitude $V_\mathrm{RF}$ as shown in \cref{fig:pulse_chain_time_sweep}c. 
A variable delay $\tau$ is introduced between the end of the RF pulse and the onset of the DC pulse; negative delay $\tau<0$ corresponds to temporal overlap of the RF and DC pulses.
We study both the $\tau \ge 0$ and $\tau<0$ regimes of RF-assisted switching.  
The phase of the RF pulse in our measurements is random (Supplementary Note 1 and Figure~S1).

For these measurements, we first select the values of $V_\mathrm{DC}$ and $t_\mathrm{DC}$ that give $\mathcal{P} = 0.5$ in the absence of an RF pulse as determined from the type of data in \cref{fig:pulse_chain_time_sweep}b for each device.
We then add an RF pulse with $t_\mathrm{RF} > t_\mathrm{DC}$ before the DC pulse and measure the increase in the switching probability $\Delta \mathcal{P}$ above the $\mathcal{P} = 0.5$ baseline.

\Cref{fig:Switching_multiple_devices} shows $\Delta \mathcal{P}$ measured as a function of RF power $P_\mathrm{RF}$ and delay $\tau$ for four device diameters. 
In these measurements, $f_\mathrm{RF}=0.1$\,GHz, $t_\mathrm{RF}=30$\,ns, and $P_\mathrm{RF}$ in dBm is defined as the power dissipated by the sample during the RF pulse:
\begin{equation}\label{eq:Powercalc}
    P_{\text{RF}}=10 \cdot \log_{10} \left( \frac{V_{\text{RF}}^2}{R_{\text{P}}}\cdot \frac{1}{1\, \text{mW}} \right),
\end{equation}  
where $R_P$ is the P state resistance (Supplementary Note~1).

\begin{figure*}[htbp]
\centering
 \includegraphics[width= 1.0\textwidth]{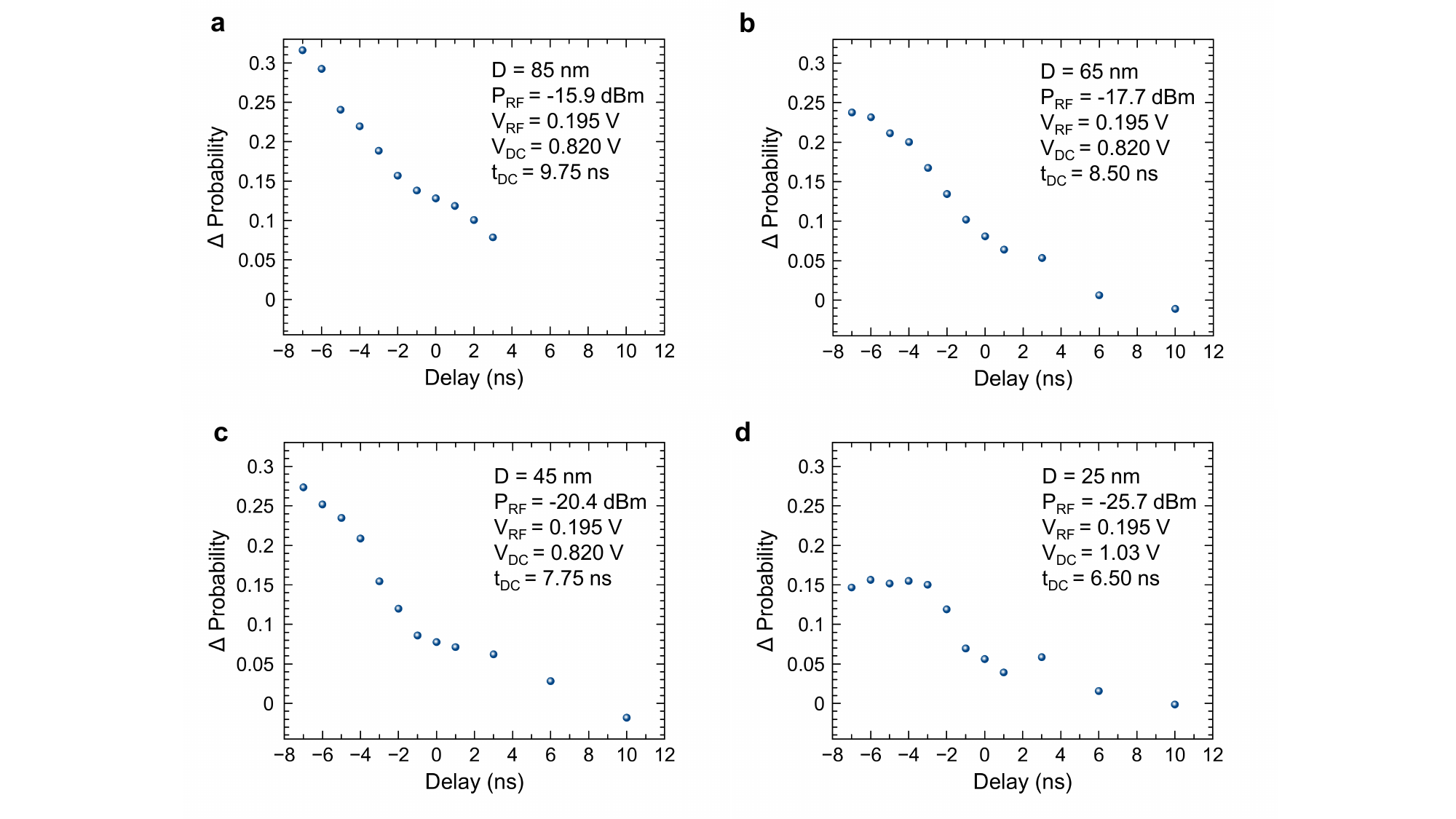}%
 \caption{\textbf{RF-assisted increase in switching probability $\Delta \mathcal{P}$ as a function of the RF-DC delay $\tau$ at $V_\mathrm{RF}=0.195$\,V.}  
  Four p-MTJ diameters are studied: \textbf{a} $D=85$\,nm. \textbf{b} $D=65$\,nm. \textbf{c} $D=45$\,nm. \textbf{d} $D=25$\,nm.
\label{fig:Max_power_switching_multiple_devices}}%
\end{figure*}
\begin{figure}[htbp]
\centering
 \includegraphics[width= 0.50\textwidth]{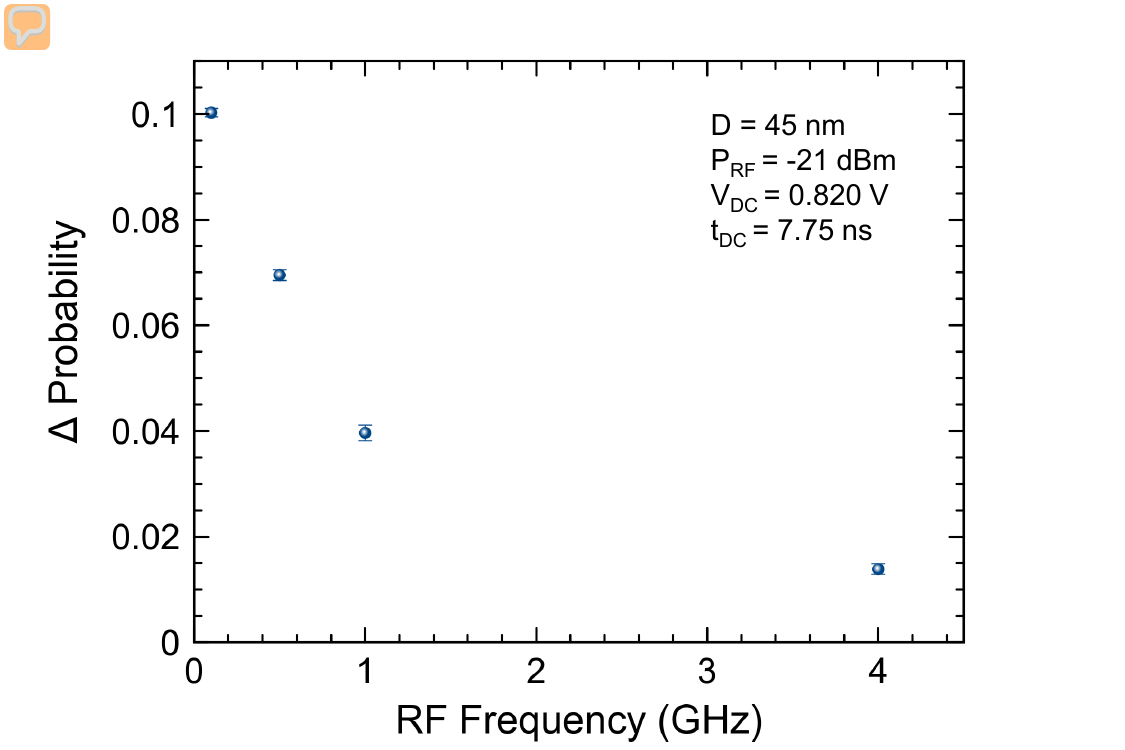}%
 \caption{\textbf{RF-assisted increase in switching probability $\Delta \mathcal{P}$ as a function of RF frequency for $\tau = -2$\,ns.} Error bars are described in Supplementary Note 1.
\label{fig:switching_vs_energy_diff_freq}}%
\end{figure}
We also study the effect of time delay between the RF and DC write pulses on $\Delta \mathcal{P}$.
\Cref{fig:Max_power_switching_multiple_devices} shows the dependence of $\Delta \mathcal{P}$ on $\tau$ at $V_\mathrm{RF} = 0.195$\,V for four MTJ diameters. $\Delta \mathcal{P}$ monotonically increases with decreasing $\tau$ and is generally positive in both the $\tau \ge 0$ and $\tau<0$ regimes. For $\tau = -7$\,ns, $\Delta \mathcal{P}$ is 0.31, 0.24, 0.27 and 0.15 for the  $D=85$, 65, 45, and 25\,nm devices, respectively. While at $\tau = 0$, $\Delta \mathcal{P}$ is 0.13, 0.08, 0.08 and 0.06 for the  $D=85$, 65, 45, and 25\,nm devices, respectively.

\Cref{fig:switching_vs_energy_diff_freq} shows the dependence of $\Delta \mathcal{P}$ on $f_\mathrm{RF}$ for the $D=45$\,nm device at $P_\mathrm{RF} = -21$\,dBm (see Supplementary Note 1 and Table~S1) and $\tau = - 2$\,ns. These data, taken at frequencies well below the FL FMR frequency (12\,GHz, Supplementary Note 2 and Figure~S3), reveal that RF-assisted switching becomes more efficient with decreasing RF frequency. A qualitatively similar dependence of $\Delta \mathcal{P}$ on $f_\mathrm{RF}$ is observed in the $\tau \ge 0$ regime.\\
\\
\textbf{\noindent\myfont Theoretical results}
\begin{figure*}
 \includegraphics[width= 0.85\textwidth]{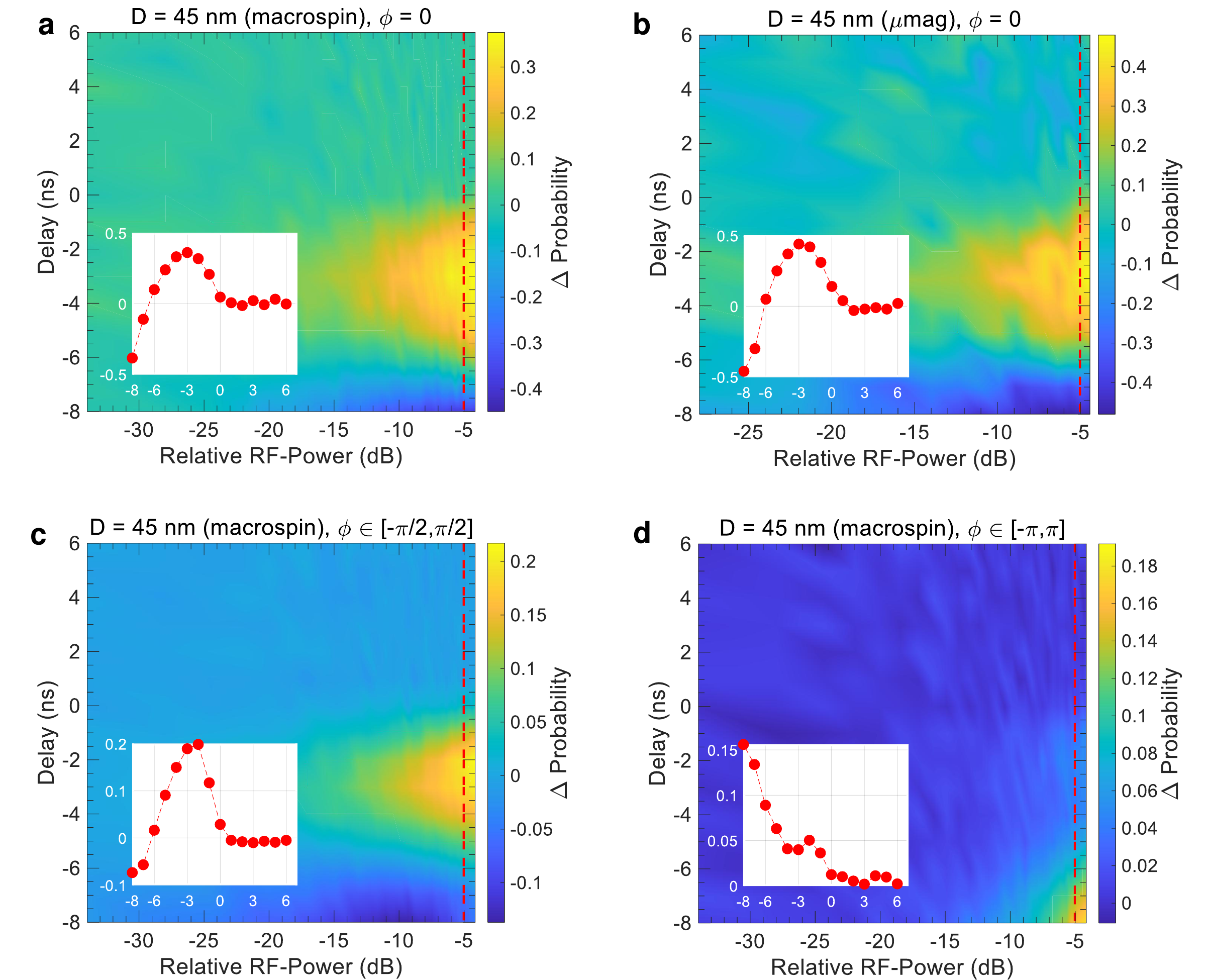}%
 \caption{{\textbf{Simulations of $\Delta \mathcal{P}$ as a function of normalized RF power $\tilde{P}_\mathrm{RF}$ and RF-DC delay $\tau$.} 
 $\Delta \mathcal{P}$ is represented by the color scale, and the insets show $\Delta \mathcal{P}$ versus $\tau$ at $\tilde{P}_\mathrm{RF}=-5$\,dB. 
 In these simulations, $f_\mathrm{RF}=0.1$\,GHz and $t_\mathrm{RF}=30$\,ns.   
 \textbf{a} Macrospin simulations with the RF phase constant $\phi = 0$.
 \textbf{b} Micromagnetic simulations with $\phi = 0$.
 \textbf{c} Macrospin simulations with $\phi$ uniformly distributed over $[-\pi/2,\pi/2]$. 
 \textbf{d} Macrospin simulations with $\phi$ uniformly distributed over $[-\pi,\pi]$.}
\label{fig:Fig_th_mltp_Prob}}%
 \end{figure*}
In order to understand the mechanisms of the experimentally observed RF-assisted switching, we made room-temperature ($T = 300$\,K) numerical macrospin and micromagnetic simulations of the switching process. 
Simulations were made for all the  MTJ diameters considered in the experiments (Supplementary Note 3).
Here we show the results for the device with $D = 45$\,nm. 
Simulations of an ensemble of multiple replicas were performed to compute the switching probability. 
The magnetization dynamics of each replica were computed using both micromagnetic and macrospin solvers (see Supplementary Note 3). 
For the micromagnetic analysis, $10^2$ replicas were simulated, while for the macrospin simulations $10^4$ replicas were used. Both the micromagnetic and the macrospin simulations were performed using Magnetization Geometrical Integration Code (MaGICo, available at \url{http://wpage.unina.it/mdaquino/index_file/MaGICo.html}).
The switching probability was then determined by counting the number of replicas that showed P to AP switching.\\
\indent For these simulations, we first select the values of $V_\mathrm{DC}$ and $t_\mathrm{DC}$ that produce $\mathcal{P} = 0.5$ in the absence of an RF pulse for each device.
We then add an RF pulse preceding the DC pulse and calculate the increase in the switching probability $\Delta \mathcal{P}$ above the $\mathcal{P} = 0.5$ baseline.
\Cref{fig:Fig_th_mltp_Prob} shows the simulated $\Delta \mathcal{P}$ due to the RF pulse assist versus normalized RF power $\tilde{P}_\mathrm{RF}$ and delay $\tau$ between the RF and DC write pulses. 
The RF power  is normalized to the minimal theoretical power required to switch the magnetic state with a DC current in the absence of thermal fluctuations ($T = 0$\,K) \cite{serpico2015_2}. 
This minimal power is analytically calculated in the macrospin approximation for the values of the FL dimensions, the uniaxial magnetocrystalline anisotropy constant $K_\mathrm{AN}$, saturation magnetization $M_\mathrm{s}$, and Gilbert damping constant $\alpha$  given in Table I of Methods.\\
\begin{figure*}
\centering
 \includegraphics[width= 1.0\textwidth]{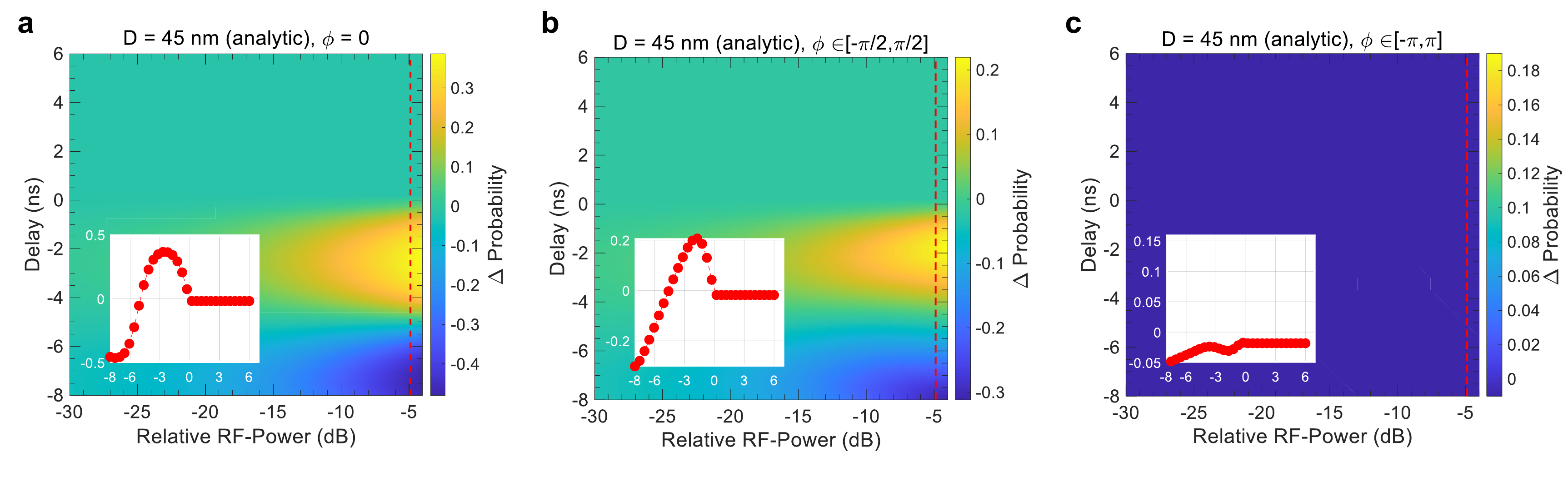}%
 \caption{\textbf{Analytic model predictions of $\Delta \mathcal{P}$ as a function of normalized RF power $\tilde{P}_\mathrm{RF}$ and RF-DC delay $\tau$.} 
 \textbf{a} $\phi = 0$.
 \textbf{b} $\phi$ uniformly distributed over $[-\pi/2,\pi/2]$. 
 \textbf{c} $\phi$ uniformly distributed over $[-\pi,\pi]$.
\label{fig:prob_ph}}%
 \end{figure*}
\indent \Cref{fig:Fig_th_mltp_Prob}a shows macrospin $\Delta \mathcal{P}$ as a function of $\tilde{P}_\mathrm{RF}$ and $\tau$ when the phase constant $\phi$ of the RF voltage $\sqrt{2} V_\mathrm{RF} \cos(\omega_\mathrm{RF} t + \phi)$ is set to zero. Notice that  $t=0$ corresponds to the time instant where the RF pulse is applied and $\omega_\mathrm{RF}=2\pi f_\mathrm{RF}$.
These results are significantly different from the experiment in \cref{fig:Switching_multiple_devices}c and \cref{fig:Max_power_switching_multiple_devices}c.
This is most clearly seen from the comparison of the inset in \cref{fig:Fig_th_mltp_Prob}a showing $\Delta \mathcal{P}$ versus $\tau$ at $\tilde{P}_\mathrm{RF} = -5$\,dB and the experimental data in \cref{fig:Max_power_switching_multiple_devices}c.
Indeed, the measured $\Delta \mathcal{P}$ is a monotonically decreasing function of $\tau$, while the simulated $\Delta \mathcal{P}(\tau)$ has a peak at $\tau = -3$\,ns. 
The micromagnetic results in \cref{fig:Fig_th_mltp_Prob}b are similar to the macrospin with a slightly enhanced $\Delta \mathcal{P}$.
The number of replicas in the macrospin simulations is large enough to yield a $\sim$1\,\% error in $\Delta \mathcal{P}$ while the micromagnetic results are precise within $\sim$10\,\% (Supplementary Note 3).\\
\indent The discrepancy with the experiment is expected because the simulations assume a fixed phase $\phi = 0$, while the measurements average over random phases.
To address this discrepancy, we performed macrospin simulations with $\phi$ uniformly distributed over $[-\pi/2,\pi/2]$ (\cref{fig:Fig_th_mltp_Prob}c) and $[-\pi,\pi]$ (\cref{fig:Fig_th_mltp_Prob}d) intervals, the latter directly mimicking the experimental conditions. The inset in \cref{fig:Fig_th_mltp_Prob}d is qualitatively similar to the corresponding experimental results in \cref{fig:Max_power_switching_multiple_devices}c, although quantitative discrepancies remain. Similar results were obtained in micromagnetic simulations and  for devices with other studied MTJ diameters (Supplementary Note 3 and Figure~S4).\\
\indent The numerically calculated dependence of $\Delta \mathcal{P}$ on $\tilde{P}_\mathrm{RF}$ and $\tau$ can be further elucidated by analytical modeling in the framework of damping-switching dynamics \cite{Serpico2008,dAquino2020}. 
This analytical model makes five assumptions:\\
(i) macrospin approximation.\\
(ii) prior to the RF+DC waveform, the distribution of replicas is assumed to be described by the Néel-Brown model \cite{Brown1963} in the high-energy-barrier case.\\ 
(iii) magnetization dynamics is governed by the deterministic Landau-Lifshitz-Gilbert (LLG) equation.\\ 
(iv) for $\tau>0$, the RF pulse does not affect the switching probability.\\
(v) for $\tau<0$, the RF+DC waveform is replaced by an effective DC pulse of duration $t_\mathrm{DC}$ and amplitude $\overline{V}_\mathrm{ST}$:
\begin{widetext}
\begin{equation}\label{eq:volt_eff}
\overline{V}_\mathrm{ST} = V_\mathrm{DC} + \sqrt{2} V_\mathrm{RF}\,\frac{|\tau|}{t_\mathrm{DC}}\mathrm{sinc}\left(\frac{\omega_\mathrm{RF}\tau}{2}\right)\cos\left(\omega_\mathrm{RF}\left(t_\mathrm{RF}+\frac{\tau}{2}\right)+\phi\right)\,,
\end{equation}
\end{widetext}
which is $V_\mathrm{DC}$ plus the average RF voltage over the duration of the DC pulse, with $t_\mathrm{RF}$ the duration of the RF pulse. 
The theoretical justification of these five assumptions is discussed in Supplementary Note 3.
Then, the magnetization switching dynamics is a damping-switching process in a uniaxial, bistable magnet between the stable states $m_{z0}$ and $m_{z1}$, where $m_{z0}$ ($m_{z1}$) belongs to the region of stability of $m_z = 1$ ($m_z = -1$). 
The magnetization trajectory from  $m_{z0}$ to $m_{z1}$ under the action of $\overline{V}_\mathrm{ST}$ can be expressed in closed form \cite{dAquino2020}, which can be used to map the ensemble of initial conditions to the ensemble of final states. Therefore, the stochastic nature of the magnetization process is included only in the statistical distribution of the initial conditions.
The FL switches if $m_z$ changes sign compared to $m_{z0}$ after the write pulse. Starting with the initial distribution with $m_{z0} > 0$, the switching probability can be computed by integrating the PDF over the range $m_z\in[-1,0]$ that defines the well of the switched magnetic state $m_z = -1$. 
This PDF solution is described in Supplementary Note 3.
The switching probability $\mathcal{P}$ for $t' = t-(t_\mathrm{RF}+\tau)>0$ is given by:

\begin{equation}\label{eq:prob_3}
\begin{aligned}
    \mathcal{P}(t') =\frac{\mathrm{erfi}\left(\sqrt{\frac{k_\mathrm{eff}}{2\tilde{T}}}\mu_z^{-1}(0,t',\overline{H})\right)}{\mathrm{erfi}\left(\sqrt{\frac{k_\mathrm{eff}}{2\tilde{T}}}\right)}\,.
    \end{aligned}
\end{equation}
The quantity $\overline{H}$ is the effective DC voltage pulse amplitude normalized to $V_\mathrm{c0}$, which is the critical DC voltage for switching at $T = 0$. The quantity $\mu_z^{-1}(0,t',\overline{H})$ is the initial condition $m_{z0}$ that by the deterministic dynamics under the action of the normalized voltage $\overline{H}$ for a time period of duration $t'$ leads to the final state $m_{z1}=0$. The quantity $k_\mathrm{eff}/\tilde{T}$ is the ratio between the normalized effective anisotropy and the normalized temperature. They are given by: $k_\mathrm{eff}=2K_\mathrm{AN}/(\mu_0M_s^2)-D_\mathrm{z}+D_\perp$, and $\tilde{T} = k_\mathrm{B}T/(\mu_0M_s^2V)$. Here, $K_\mathrm{AN}$ is the uniaxial magnetocrystalline anisotropy, $D_\mathrm{z}$ and $D_\mathrm{\perp}$ are the demagnetizing factors along the symmetry axis and the perpendicular direction $(2D_\perp +D_z = 1)$, $k_\mathrm{B}$ is the Boltzmann constant, $\mu_0$ is the vacuum magnetic permeability, $M_s$ is the saturation magnetization, and $V$ is the volume of the free layer. All the quantities are expressed in SI units.
Taking into account that PDF depends on $\phi$ via $\overline{H}$, the switching probability for $\phi$ uniformly distributed over the experimentally relevant interval $[-\pi,\pi]$ is:
\begin{equation}\label{eq:prob_4}
    \mathcal{P}(t') =\frac{\int_0^{2\pi}\mathrm{erfi}\left(\sqrt{\frac{k_\mathrm{eff}}{2\tilde{T}}}\mu_z^{-1}(0,t',\overline{H})\right)d\phi}{2\pi\,\mathrm{erfi}\left(\sqrt{\frac{k_\mathrm{eff}}{2\tilde{T}}}\right)}\,.
\end{equation}
In \cref{fig:prob_ph}, we use \cref{eq:prob_3} and \cref{eq:prob_4} to compute $\Delta \mathcal{P}$ for the same parameters as numerical results in \cref{fig:Fig_th_mltp_Prob}. 
In the case of $\phi = 0$ and $\phi\in[-\pi/2,\pi/2]$ (\cref{fig:prob_ph}a,b), the agreement with the numerical results is quantitative, but its quality becomes worse for large values of $P_\mathrm{RF}$ and $\tau<0$, where the analytical results underestimate $\Delta \mathcal{P}$. 
Since the effective write pulse amplitude $\overline{V}_\mathrm{ST}$ is an oscillatory function of $\tau$, $P_\mathrm{RF}$ oscillates between positive and negative values for $\tau<0$.

In the case of the fully random phase $\phi$ distributed in $\phi\in[-\pi,\pi]$, analytical calculations in \cref{fig:prob_ph}c do not predict an increase of $\mathcal{P}$, while numerical simulations in \cref{fig:Fig_th_mltp_Prob}d show $\Delta \mathcal{P}>0$ for large values of $P_\mathrm{RF}$ and $\tau<0$.
This is because $\mathcal{P}(\phi)$ in \cref{eq:prob_3} is a periodic function of $\phi$, and the enhancement of $\mathcal{P}$ achieved for a given $\phi^*$ is approximately canceled by the reduction of $\mathcal{P}$ for the phase $\phi^*+\pi$. 
Although this cancellation is not exact, $\Delta \mathcal{P}$ resulting from the averaging in $\phi$ is small $\sim 0.01$.

In numerical simulations in \cref{fig:Fig_th_mltp_Prob}d with sufficiently high $\tilde{P}_\mathrm{RF}$ and $\tau<0$, $\Delta \mathcal{P}$ is not negligibly small because a larger number of switching events with respect to those predicted by the analytical model in \cref{fig:prob_ph}c occur in the time interval where the RF pulse adds constructively to the DC pulse voltage. This fact is not surprising since the assumption (v) of the analytical model does not hold anymore (Supplementary Note 3).
As a result, numerical simulations produce higher $\Delta \mathcal{P}$ than the analytic model.\\

\begin{large}
\noindent \textbf{\myfont Discussion}
\end{large}

\indent \Cref{fig:Switching_multiple_devices} reveals that RF-assisted switching enhancement is observed for both positive and negative delays $\tau$ between the RF and DC pulses. 
The enhancement for $\tau < 0$ can be expected because the overlap between the RF and DC pulses can result in the net applied voltage exceeding $V_\mathrm{DC}$ in an interval of the combined RF+DC waveform. 
However, the RF-enhanced switching in the $\tau \ge 0$ regime is surprising. 
This latter regime is also important from a practical point of view because the combined RF+DC waveform voltage never exceeds $V_\mathrm{DC}$, leading to enhanced switching without extra stress on the junction. 
From another angle, introduction of an RF pulse allows one to reduce the DC pulse duration $t_\mathrm{DC}$ for a targeted switching probability compared to the DC-only protocol. 
Since longer DC write pulses are known to stress the MTJ barrier oxide and reduce device endurance \cite{amara2012, gapihan2012, wang_different_2014}, the RF+DC switching protocol in the $\tau \ge 0$ regime studied here is expected to enhance STT-MRAM endurance. 
Prior studies of STT-MRAM reported a nearly order-of-magnitude improvement in the MgO write cycles before the dielectric breakdown when $t_\mathrm{DC}$ was reduced from $10$\,ns to $5$\,ns \cite{wang_different_2014}.

The maximum RF pulse amplitude used in this study, $V_\mathrm{RF} = 0.195$\,V, is much smaller than the DC write pulse amplitude $\sim 1$\,V. 
Therefore, in the $\tau \ge 0$ regime, this small-amplitude RF assist does not diminish device endurance and only marginally increases the write energy while increasing the switching probability.
This makes RF assist at $\tau = 0$ a viable strategy for an improved STT-MRAM write protocol. 
For example, addition of a small-amplitude RF pulse at $\tau = 0$ in \cref{fig:Max_power_switching_multiple_devices}a, leads to a modest 17\,\% increase of the write energy while giving a 13\,\% increase in the switching probability without affecting the device endurance.
Our calculations of the energy per write (Supplementary Note 1 and Figure~S2) identify regimes in which the RF+DC and DC protocols have equal write energies, while $t_{\mathrm{DC}}$ is shorter in the RF+DC protocol. 
Figure~S2 shows an example where the addition of a small-amplitude RF pulse reduces $t_{\mathrm{DC}}$ while maintaining the same total write energy for $\tau = -7$\,ns. 
A comprehensive follow-up study of STT switching as a function of RF and DC pulse parameters is required to determine whether the RF+DC protocol can achieve higher energy efficiency than the DC-only write protocol. 
The present study focuses on regimes $\tau \ge 0$ where the RF+DC protocol improves the switching probability without compromising STT-MRAM endurance.\\
\indent\Cref{fig:Switching_multiple_devices} also demonstrates that the small-amplitude RF pulse increases the switching probability for all device sizes down to $D = 25$\,nm. 
This is promising for scalability of the RF+DC write. 
While the observed enhancement of switching probability $\Delta \mathcal{P}$ at $V_\mathrm{RF} = 0.195$\,V is smaller for the 25\,nm device compared to the 85\,nm device, the RF power dissipated by the 25\,nm device at this RF assist voltage is 10\,dB lower due to the higher resistance of the smaller MTJ. 
The effectiveness of the RF assist for all device sizes indicates that micromagnetic effects expected to be more pronounced in the larger devices do not play a crucial role in the observed effect. 
This is corroborated by our theoretical results in \cref{fig:Fig_th_mltp_Prob} revealing similar impact of the RF assist in the macrospin and micromagnetic approximations.\\
\indent In principle, ohmic heating induced by the RF pulse can contribute to the enhancement of switching probability $\Delta \mathcal{P}$.  
However, because time-averaged ohmic heating is frequency independent, MTJ heating averaged over the RF pulse duration cannot account for the strong frequency dependence seen in \cref{fig:switching_vs_energy_diff_freq}.
Circuit calibrations used in our experiments ensured uniform RF pulse power across all frequencies (See Supplementary Note 1 and Table~S1). 
If time-averaged MTJ temperature increase induced by the RF pulse were the dominant mechanism, the observed switching enhancement would be comparable across all frequencies at constant power. 
The strong frequency dependence of the RF assist demonstrated in \cref{fig:switching_vs_energy_diff_freq} therefore eliminates time-averaged ohmic heating of the MTJ by the RF pulse as the primary mechanism of the observed RF-induced switching enhancement \cite{rehm2021}.

Our theoretical predictions of RF-assisted switching in \cref{fig:Fig_th_mltp_Prob}d are qualitatively similar to the experimental results in \cref{fig:Switching_multiple_devices}c and \cref{fig:Max_power_switching_multiple_devices}c. 
The theory captures the threshold character of the effect in the RF power and the observed decrease in the switching probability with the delay time $\tau$ between the RF and DC pulses.  
However, there are two quantitative differences. 
First, the theoretical threshold power is higher than the experimental one. 
Second, theoretical results shown in the insets of \cref{fig:Fig_th_mltp_Prob} suggest that the effect vanishes for delay times $\tau$ exceeding $\sim 1$\,ns, while the observed effect in \cref{fig:Max_power_switching_multiple_devices} persists up to $\tau \approx 6$\,ns.\\
\indent The theoretically predicted $\sim 1$\,ns time scale is reasonable because it is similar to the characteristic magnetic energy relaxation time of the FL $\tau_E = 1/\left(4 \pi \alpha f_\mathrm{FMR} \right) = 0.66$\,ns, where $\alpha = 0.0097$ and $f_\mathrm{FMR} = 12$\,GHz as determined by spin torque FMR measurements (Supplementary Note 2 and Figure~S3).  
For $\tau \gg \tau_E$, all magnetic energy added to the FL magnetization by the RF pulse is expected to dissipate away, and thus magnetization dynamics driven by the DC pulse is expected to be unaffected by the RF pulse. 
The observation of RF assist effects for $\tau \gg \tau_E$ suggests that, in addition to FMR dynamics, the FL also supports significantly slower magnetic dynamics. \\
\indent Signatures of slow magnetic dynamics have been previously observed in a different experiment \cite{safranski2019interface} and were attributed to regions at the FL/MgO interface with reduced magnetization and weak exchange coupling. 
Our data in \cref{fig:Max_power_switching_multiple_devices} provide additional support for this hypothesis of slow dynamics in the FL of a p-MTJ. Moreover, the greater enhancement of switching probability $\Delta \mathcal{P}$ at lower RF frequencies as shown in \cref{fig:switching_vs_energy_diff_freq} suggests these dynamics are more efficiently excited at lower frequencies.
This effect may be rooted in the structural, chemical, and magnetic inhomogeneity of the ultrathin CoFeB FL found in MTJ devices \cite{barsukov2015magnetic}.
Electrical transport, FMR and magnetometry studies on ultrathin films and MTJ nanopillars based on Ta/CoFeB/MgO multilayers suggest that iron-based magnetic oxides form at the CoFeB/MgO interface \cite{barsukov2015magnetic, barsukov2014}.   
These slow magnetic dynamics attributed to the FL inhomogeneity could contribute to the enhancement of switching probability $\Delta \mathcal{P}$ for $\tau \geq 0$ but are beyond the scope of the theoretical framework of this paper.\\
\indent The threshold character of the RF-induced enhancement of switching in \cref{fig:Switching_multiple_devices} suggests the importance of nonlinear magneto-dynamics \cite{chen2017} for the RF assist. 
Both the threshold character of the RF assist in the RF power and the frequency dependence of its efficiency in \cref{fig:switching_vs_energy_diff_freq} are consistent with the prediction of magnetization reversal driven by low-dimensional chaos induced by the RF drive \cite{montoya2019magnetization}. 
However, this mechanism is expected to be inactive in perfectly axially symmetric cylindrical structures with uniaxial magnetic anisotropy. This is because the dynamic system is effectively two-dimensional with one dimension represented by the magnetization polar angle and the other by the RF drive phase. According to the Poincaré – Bendixson theorem \cite{ciesielski2012}, such a system does not exhibit chaotic dynamics.

However, real nominally circular MTJ devices are known to deviate from this ideal picture and exhibit a weaker random in-plane uniaxial anisotropy in addition to the dominant out-of-plane uniaxial anisotropy \cite{hirayama2015, mazraati2016}. 
This anisotropy can arise from both the MTJ shape imperfections and from crystallographic anisotropy of the randomly oriented FL grains. 
In this case, both the polar and azimuthal angles of the magnetization enter the system's Lagrange function, and the RF drive makes the system effectively three-dimensional \cite{serpico2015, suarez2017, ferona2017}. 
In such a system, RF-driven chaotic dynamics is expected above a threshold amplitude of the drive, and the threshold magnitude decreases with deceasing frequency. 
This RF-driven chaos mechanism can contribute to the observed RF assist and reduce the threshold for the onset of RF-induced switching over that theoretically predicted for ideal cylindrically-symmetric devices in \cref{fig:Fig_th_mltp_Prob}.

A potential mechanism of switching probability enhancement not included in the present theoretical description is the dynamics of the FL temperature and its coupling with the magnetization switching dynamics. 
An RF current applied to the MTJ produces a temperature increase that has a time-independent component that also does not depend on $f_\mathrm{RF}$ and an oscillatory component at the frequency $2 f_\mathrm{RF}$ with a frequency-dependent amplitude approximately given by $T_0/\sqrt{1+(4 \pi \tau_\mathrm{\theta} f_\mathrm{RF})^2}$, where $\tau_\mathrm{\theta}$ is the thermal time constant of the system and $T_0$ is the temperature oscillation amplitude in the low frequency limit \cite{sullivan_1968, Rogalski2010}. 
If we assume that the frequency dependence of the RF assisted switching in \cref{fig:switching_vs_energy_diff_freq} is linked to these time-dependent temperature oscillations, we can fit the data in \cref{fig:switching_vs_energy_diff_freq} to $\Delta \mathcal{P}_0/\sqrt{1+(4 \pi \tau_\mathrm{\theta} f_\mathrm{RF})^2}$ in order to extract $\tau_\mathrm{\theta}$,  where $\Delta\mathcal{P}_0$ is the switching probability enhancement in the low frequency limit. Such a fitting procedure gives $\tau_\mathrm{\theta}=0.18 \pm 0.01$\,ns.

Such a small thermal time constant is an order of magnitude below previously reported values of $\tau_\mathrm{\theta}$ for nanoscale MTJs \cite{chavent_2016}. In addition, $\tau \gg \tau_\theta$ is inconsistent with the data in \cref{fig:Max_power_switching_multiple_devices}, which shows that the effects of the RF pulse on the switching dynamics persists for nanoseconds after the end of the RF pulse. Note that the RF pulses of the magnitude and duration used in this study do not produce any measurable MTJ switching on their own without a follow-up DC pulse.

We, therefore, conclude that while a contribution of Ohmic heating to the observed RF assist cannot be completely excluded, it cannot be a simple process governed by a single time constant. Investigating a more complex mechanism would require careful measurements of the thermal time constants together with the development of a theory describing coupled thermal and magnetization dynamics. This analysis is beyond the scope of the present work and will be addressed in future studies.
\newpage
\begin{large}
\noindent \textbf{\myfont Methods}
\end{large}

\noindent \textbf{\myfont Sample description}
The bottom electrode/ SAF/ MgO/ CoFeB(2.21\,nm)/ top electrode multilayer stack was deposited by magnetron sputtering and patterned into circular MTJs using e-beam lithography and ion mill etching. 
Saturation magnetization of the CoFeB FL was determined by vibrating sample magnetometry, while its out-of-plane uniaxial magnetic anisotropy and Gilbert damping were determined by ST-FMR measurements (Supplementary Note 2) as summarized in \cref{tab1}.

\begin{table}[ht]
    \caption{Material parameters of the MTJ CoFeB free layer: thickness, saturation magnetization $M_\mathrm{s}$, uniaxial magnetocrystalline anisotropy constant $K_{AN}$, Gilbert damping constant $\alpha$.\\}\label{tab1}
    \centering
    \begin{tabular}{ c c c }
        \hline
        Quantity & Value & Unit\\
        \hline
        FL thickness & $2.21$ & nm \\
        $M_s$ & $1039$ & $\mathrm{kA/m}$ \\
        $K_\mathrm{AN}$ & $8.6 \times 10^5$ & $\mathrm{J/m^3}$\\
        $\alpha$ & $0.0097$ \\
        \hline
    \end{tabular}
\end{table}

\noindent \textbf{\myfont Spin torque ferromagnetic resonance} 
We use ST-FMR to measure the frequency and linewidth of the FL spin wave modes as a function of out-of-plane magnetic field, and extract the FL Gilbert damping constant $\alpha$ and the uniaxial magnetocrystalline anisotropy constant $K_{AN}$ at room temperature from these data \cite{gonccalves2013spin, safranski2016, barsukov2019}.  The values presented in \cref{tab1} are an ensemble average of multiple devices.  
In these measurements, a microwave voltage is applied to an MTJ via the RF port of a bias tee.  
The applied voltage induces an RF spin transfer torque that excites spin waves in the FL and leads to oscillations of the sample resistance at the drive frequency due to tunneling magneto-resistance of the MTJ. 
These resistance oscillations mix with the RF current through the MTJ and generate a rectified DC voltage, which is measured by a lock-in amplifier through the DC port of the bias tee. 
The magnetic field modulation technique is employed to improve the signal-to-noise ratio \cite{gonccalves2013spin}.\\ 

\noindent \textbf{\myfont RF-assisted switching probability measurements}
Current-induced switching of the MTJ FL is observed by monitoring the change in the MTJ resistance in a circuit where the MTJ forms a voltage divider with a fixed resistor.
We bias the voltage divider by a low-level DC voltage and use a 16-bit 2 MHz data acquisition board (DAQ) to measure time-dependent voltage across the MTJ.
To switch the FL, we apply a DC voltage pulse preceded by an RF voltage pulse to the MTJ. 
Details of the circuit and the waveform protocol are given in Supplementary Note 1. 
Following the RF+DC write waveform, we use DAQ to measure the resultant resistance state of the MTJ, which registers either switching or non-switching.
A lower-amplitude, longer-duration DC reset pulse of the opposite polarity to the DC write pulse is then used to return the device to its initial state, and the DAQ is used to confirm the successful reset.  
The RF pulse is generated by using a broadband microwave mixer connected to a microwave generator controlling the RF frequency and a square pulse generator controlling the output RF pulse duration. The amplitude of the RF pulse is controlled by both generators.
All three pulse generators used in the measurements are synchronized with each other and the DAQ as described in Supplementary Note 1. 
The microwave generator operates in continuous wave mode, resulting in a random phase $\phi$ of the RF pulse.  
The delay $\tau$ between the RF and DC pulses is controlled through a combination of the cable length and a programmable internal delay of the DC write pulse generator. 
The switching probability $\mathcal{P}$  is determined as a ratio of the number of successful switching events to the total number of switching attempts.
The voltage levels of all RF+DC waveforms used in the measurements are characterized using a high-bandwidth real-time oscilloscope.\\

\begin{large}
\noindent \textbf{\myfont Data availability}
\end{large}

\noindent All data generated or analyzed during this study are included in this published article and are available from the corresponding author upon reasonable request.

\bigskip

\begin{large}
\noindent \textbf{\myfont Acknowledgements}
\end{large}

\noindent I.N.K. and P.M.B. thank Jordan Katine for stimulating discussions.\\

\noindent This work was primarily supported by Western Digital Technologies (grant number: not applicable).  
Funding from the National Science Foundation via awards ECCS-2213690 and DMREF-2324203 as well as from the University of California National Laboratory Fees Research Program (grant number: not applicable) is also acknowledged.
The authors thank the Eddleman Quantum Institute (EQI) for partial support of this work (grant number: not applicable).
S.P gratefully acknowledges financial support for this publication by the Fulbright Visiting Scholar Program, which is sponsored by the U.S. Department of State and the Italian Ministry of Foreign Affairs and administered by the U.S. - Italy Fulbright Commission (grant number: not applicable).  The contents of this publication are solely the responsibility of the author and do not necessarily represent the official views of the Fulbright Program, the Government of the United States, or the U.S. - Italy Fulbright Commission.\\

\begin{large}
\noindent \textbf{\myfont Author contributions}
\end{large}

\noindent P.M.B., W.J. and C.D. made the samples. M.H. and I.N.K. designed the experiment. M.H. made the measurements. S.P., M.d'A. and C.S. developed the theoretical model. S.P. made numerical simulations. P.M.B and I.N.K. conceived the idea and
supervised the project.\\

\begin{large}
\noindent \textbf{\myfont Competing interests}
\end{large}

\noindent The authors declare no competing interests.

\bibliography{refer}

@article{safranski2019interface,
	title = {Interface moment dynamics and its contribution to spin-transfer torque switching process in magnetic tunnel junctions},
	volume = {100},
	issn = {2469-9950, 2469-9969},
	url = {https://link.aps.org/doi/10.1103/PhysRevB.100.014435},
	doi = {10.1103/PhysRevB.100.014435},
	language = {en},
	number = {1},
	urldate = {2025-08-29},
	journal = {Physical Review B},
	author = {Safranski, Christopher and Sun, Jonathan Z.},
	month = jul,
	year = {2019},
	pages = {014435},
}

@article{barsukov2015magnetic,
	title = {Magnetic phase transitions in {Ta}/{CoFeB}/{MgO} multilayers},
	volume = {106},
	issn = {0003-6951, 1077-3118},
	url = {https://pubs.aip.org/apl/article/106/19/192407/27672/Magnetic-phase-transitions-in-Ta-CoFeB-MgO},
	doi = {10.1063/1.4921306},
	language = {en},
	number = {19},
	urldate = {2025-08-29},
	journal = {Applied Physics Letters},
	author = {Barsukov, I. and Fu, Yu and Safranski, C. and Chen, Y.-J. and Youngblood, B. and Gonçalves, A. M. and Spasova, M. and Farle, M. and Katine, J. A. and Kuo, C. C. and Krivorotov, I. N.},
	month = may,
	year = {2015},
	pages = {192407},
}

@article{gonccalves2013spin,
	title = {Spin torque ferromagnetic resonance with magnetic field modulation},
	volume = {103},
	issn = {0003-6951, 1077-3118},
	url = {https://pubs.aip.org/apl/article/103/17/172406/26456/Spin-torque-ferromagnetic-resonance-with-magnetic},
	doi = {10.1063/1.4826927},
	language = {en},
	number = {17},
	urldate = {2025-08-29},
	journal = {Applied Physics Letters},
	author = {Gonçalves, A. M. and Barsukov, I. and Chen, Y.-J. and Yang, L. and Katine, J. A. and Krivorotov, I. N.},
	month = oct,
	year = {2013},
	pages = {172406},
}

@article{worledge2024spin,
	title = {Spin-transfer torque magnetoresistive random access memory technology status and future directions},
	volume = {1},
	issn = {2948-1201},
	url = {https://www.nature.com/articles/s44287-024-00111-z},
	doi = {10.1038/s44287-024-00111-z},
	language = {en},
	number = {11},
	urldate = {2025-08-29},
	journal = {Nature Reviews Electrical Engineering},
	author = {Worledge, Daniel C. and Hu, Guohan},
	month = nov,
	year = {2024},
	pages = {730--747},
}

@article{montoya2019magnetization,
	title = {Magnetization reversal driven by low dimensional chaos in a nanoscale ferromagnet},
	volume = {10},
	issn = {2041-1723},
	url = {https://www.nature.com/articles/s41467-019-08444-2},
	doi = {10.1038/s41467-019-08444-2},
	language = {en},
	number = {1},
	urldate = {2025-08-29},
	journal = {Nature Communications},
	author = {Montoya, Eric Arturo and Perna, Salvatore and Chen, Yu-Jin and Katine, Jordan A. and d’Aquino, Massimiliano and Serpico, Claudio and Krivorotov, Ilya N.},
	month = feb,
	year = {2019},
	pages = {543},
}

@article{cui2008resonant,
	title = {Resonant spin-transfer-driven switching of magnetic devices assisted by microwave current pulses},
	volume = {77},
	copyright = {http://link.aps.org/licenses/aps-default-license},
	issn = {1098-0121, 1550-235X},
	url = {https://link.aps.org/doi/10.1103/PhysRevB.77.214440},
	doi = {10.1103/PhysRevB.77.214440},
	language = {en},
	number = {21},
	urldate = {2025-08-29},
	journal = {Physical Review B},
	author = {Cui, Y.-T. and Sankey, J. C. and Wang, C. and Thadani, K. V. and Li, Z.-P. and Buhrman, R. A. and Ralph, D. C.},
	month = jun,
	year = {2008},
	pages = {214440},
}

@article{dAquino2020,
	title = {Micromagnetic study of statistical switching in magnetic tunnel junctions stabilized by perpendicular shape anisotropy},
	volume = {577},
	issn = {09214526},
	url = {https://linkinghub.elsevier.com/retrieve/pii/S0921452619306465},
	doi = {10.1016/j.physb.2019.411744},
	language = {en},
	urldate = {2025-08-29},
	journal = {Physica B: Condensed Matter},
	author = {d'Aquino, M. and Perna, S. and Serpico, C.},
	month = jan,
	year = {2020},
	pages = {411744},
}

@article{Brown1963,
	title = {Thermal {Fluctuations} of a {Single}-{Domain} {Particle}},
	volume = {130},
	copyright = {http://link.aps.org/licenses/aps-default-license},
	issn = {0031-899X},
	url = {https://link.aps.org/doi/10.1103/PhysRev.130.1677},
	doi = {10.1103/PhysRev.130.1677},
	language = {en},
	number = {5},
	urldate = {2025-08-30},
	journal = {Physical Review},
	author = {Brown, William Fuller},
	month = jun,
	year = {1963},
	pages = {1677--1686},
}

@BOOK{Serpico2008,
	author={Mayergoyz, I.D. and Bertotti, G. and Serpico, C.},
	title={Nonlinear magnetization dynamics in nanosystems},
	isbn      = {9780080443164},
    doi       = {10.1016/b978-0-08-044316-4.x0001-1},
    series={Elsevier Series in Electromagnetism},	
    year={2009},
    month     = oct,
    address		= "Amsterdam",
    Publisher={Elsevier Science},
    url = {https://linkinghub.elsevier.com/retrieve/pii/B9780080443164X00011},
}

@article{apalkov2016,
	title = {Magnetoresistive Random Access Memory},
	volume = {104},
	issn = {1558-2256},
	url = {https://ieeexplore.ieee.org/document/7555318},
	doi = {10.1109/JPROC.2016.2590142},
	pages = {1796--1830},
	number = {10},
	journal = {Proceedings of the IEEE},
	author = {Apalkov, Dmytro and Dieny, Bernard and Slaughter, J. M.},
	urldate = {2025-08-16},
	year = {2016},
}

@article{rizzo2013,
	title = {A {Fully} {Functional} 64 {Mb} {DDR3} {ST}-{MRAM} {Builton} 90 nm {CMOS} {Technology}},
	volume = {49},
	issn = {0018-9464},
	doi = {10.1109/TMAG.2013.2243133},
	number = {7},
	journal = {IEEE Transactions on Magnetics},
	author = {Rizzo, N. D. and Houssameddine, D. and Janesky, J. and Whig, R. and Mancoff, F. B. and Schneider, M. L. and DeHerrera, M. and Sun, J. J. and Nagel, K. and Deshpande, S. and Chia, H.- and Alam, S. M. and Andre, T. and Aggarwal, S. and Slaughter, J. M.},
	month = jul,
	year = {2013},
	pages = {4441--4446},
}

@inproceedings{sakhare2018,
	title = {Enablement of STT-MRAM as last level cache for the high performance computing domain at the 5 nm node},
	url = {https://ieeexplore.ieee.org/document/8614637},
	doi = {10.1109/IEDM.2018.8614637},
	urldate = {2025-08-17},
	booktitle = {2018 {IEEE} {International} {Electron} {Devices} {Meeting} ({IEDM})},
	author = {Sakhare, S. and Perumkunnil, M. and Bao, T. Huynh and Rao, S. and Kim, W. and Crotti, D. and Yasin, F. and Couet, S. and Swerts, J. and Kundu, S. and Yakimets, D. and Baert, R. and Oh, HR. and Spessot, A. and Mocuta, A. and Kar, G. Sankar and Furnemont, A.},
	month = dec,
	year = {2018},
	pages = {18.3.1--18.3.4},
}

@article{zhao2016,
	title = {Failure {Analysis} in {Magnetic} {Tunnel} {Junction} {Nanopillar} with {Interfacial} {Perpendicular} {Magnetic} {Anisotropy}},
	volume = {9},
	copyright = {http://creativecommons.org/licenses/by/3.0/},
	issn = {1996-1944},
	url = {https://www.mdpi.com/1996-1944/9/1/41},
	doi = {10.3390/ma9010041},
	language = {en},
	number = {1},
	urldate = {2025-08-18},
	journal = {Materials},
	author = {Zhao, Weisheng and Zhao, Xiaoxuan and Zhang, Boyu and Cao, Kaihua and Wang, Lezhi and Kang, Wang and Shi, Qian and Wang, Mengxing and Zhang, Yu and Wang, You and Peng, Shouzhong and Klein, Jacques-Olivier and De Barros Naviner, Lirida Alves and Ravelosona, Dafine},
	month = jan,
	year = {2016},
	pages = {41},
}

@article{perrissin2018,
	title = {A highly thermally stable sub-20 nm magnetic random-access memory based on perpendicular shape anisotropy},
	volume = {10},
	issn = {2040-3372},
	url = {https://pubs.rsc.org/en/content/articlelanding/2018/nr/c8nr01365a},
	doi = {10.1039/C8NR01365A},
	language = {en},
	number = {25},
	urldate = {2025-08-18},
	journal = {Nanoscale},
	author = {Perrissin, N. and Lequeux, S. and Strelkov, N. and Chavent, A. and Vila, L. and Buda-Prejbeanu, L. D. and Auffret, S. and Sousa, R. C. and Prejbeanu, I. L. and Dieny, B.},
	month = jul,
	year = {2018},
	pages = {12187--12195},
}

@article{watanabe2018,
	title = {Shape anisotropy revisited in single-digit nanometer magnetic tunnel junctions},
	volume = {9},
	copyright = {2018 The Author(s)},
	issn = {2041-1723},
	url = {https://www.nature.com/articles/s41467-018-03003-7},
	doi = {10.1038/s41467-018-03003-7},
	language = {En},
	number = {1},
	urldate = {2019-07-19},
	journal = {Nature Communications},
	author = {Watanabe, K. and Jinnai, B. and Fukami, S. and Sato, H. and Ohno, H.},
	month = feb,
	year = {2018},
	pages = {663},
}

@article{rowlands2011,
	title = {Deep subnanosecond spin torque switching in magnetic tunnel junctions with combined in-plane and perpendicular polarizers},
	volume = {98},
	issn = {0003-6951, 1077-3118},
	url = {http://scitation.aip.org/content/aip/journal/apl/98/10/10.1063/1.3565162},
	doi = {10.1063/1.3565162},
	number = {10},
	urldate = {2015-05-05},
	journal = {Applied Physics Letters},
	author = {Rowlands, G. E. and Rahman, T. and Katine, J. A. and Langer, J. and Lyle, A. and Zhao, H. and Alzate, J. G. and Kovalev, A. A. and Tserkovnyak, Y. and Zeng, Z. M. and Jiang, H. W. and Galatsis, K. and Huai, Y. M. and Khalili Amiri, P. and Wang, K. L. and Krivorotov, I. N. and Wang, J.-P.},
	month = mar,
	year = {2011},
	pages = {102509},
}

@article{liu2012,
	title = {Current-{Induced} {Switching} of {Perpendicularly} {Magnetized} {Magnetic} {Layers} {Using} {Spin} {Torque} from the {Spin} {Hall} {Effect}},
	volume = {109},
	url = {https://link.aps.org/doi/10.1103/PhysRevLett.109.096602},
	doi = {10.1103/PhysRevLett.109.096602},
	number = {9},
	urldate = {2025-08-18},
	journal = {Physical Review Letters},
	author = {Liu, Luqiao and Lee, O. J. and Gudmundsen, T. J. and Ralph, D. C. and Buhrman, R. A.},
	month = aug,
	year = {2012},
	pages = {096602},
}

@article{miron2011,
	title = {Perpendicular switching of a single ferromagnetic layer induced by in-plane current injection},
	volume = {476},
	copyright = {2011 Springer Nature Limited},
	issn = {1476-4687},
	url = {https://www.nature.com/articles/nature10309},
	doi = {10.1038/nature10309},
	language = {en},
	number = {7359},
	urldate = {2025-08-18},
	journal = {Nature},
	author = {Miron, Ioan Mihai and Garello, Kevin and Gaudin, Gilles and Zermatten, Pierre-Jean and Costache, Marius V. and Auffret, Stéphane and Bandiera, Sébastien and Rodmacq, Bernard and Schuhl, Alain and Gambardella, Pietro},
	month = aug,
	year = {2011},
	pages = {189--193},
}

@article{yang2024,
	title = {Field-free spin–orbit torque switching in ferromagnetic trilayers at sub-ns timescales},
	volume = {15},
	copyright = {2024 The Author(s)},
	issn = {2041-1723},
	url = {https://www.nature.com/articles/s41467-024-46113-1},
	doi = {10.1038/s41467-024-46113-1},
	language = {en},
	number = {1},
	urldate = {2025-08-18},
	journal = {Nature Communications},
	author = {Yang, Qu and Han, Donghyeon and Zhao, Shishun and Kang, Jaimin and Wang, Fei and Lee, Sung-Chul and Lei, Jiayu and Lee, Kyung-Jin and Park, Byong-Guk and Yang, Hyunsoo},
	month = feb,
	year = {2024},
	pages = {1814},
}

@article{lopez-dominguez2023,
	title = {Perspectives on field-free spin–orbit torque devices for memory and computing applications},
	volume = {133},
	issn = {0021-8979},
	url = {https://doi.org/10.1063/5.0135185},
	doi = {10.1063/5.0135185},
	number = {4},
	urldate = {2025-08-18},
	journal = {Journal of Applied Physics},
	author = {Lopez-Dominguez, Victor and Shao, Yixin and Khalili Amiri, Pedram},
	month = jan,
	year = {2023},
	pages = {040902},
}

@article{baek2019,
	title = {Limited {Stochastic} {Current} for {Energy}-{Optimized} {Switching} of {Spin}-{Transfer}-{Torque} {Magnetic} {Random}-{Access} {Memory}},
	volume = {12},
	url = {https://link.aps.org/doi/10.1103/PhysRevApplied.12.064004},
	doi = {10.1103/PhysRevApplied.12.064004},
	number = {6},
	urldate = {2025-08-19},
	journal = {Physical Review Applied},
	author = {Baek, Eunchong and Purnama, Indra and You, Chun-Yeol},
	month = dec,
	year = {2019},
	pages = {064004},
}

@article{sun2021,
	title = {Spin-transfer torque switching probability of {CoFeB}/{MgO}/{CoFeB} magnetic tunnel junctions beyond macrospin},
	volume = {104},
	url = {https://link.aps.org/doi/10.1103/PhysRevB.104.104428},
	doi = {10.1103/PhysRevB.104.104428},
	number = {10},
	urldate = {2025-08-19},
	journal = {Physical Review B},
	author = {Sun, J. Z.},
	month = sep,
	year = {2021},
	pages = {104428},
}

@article{carpentieri2010,
	title = {Spin-transfer-torque resonant switching and injection locking in the presence of a weak external microwave field for spin valves with perpendicular materials},
	volume = {82},
	url = {https://link.aps.org/doi/10.1103/PhysRevB.82.094434},
	doi = {10.1103/PhysRevB.82.094434},
	number = {9},
	urldate = {2025-08-19},
	journal = {Physical Review B},
	author = {Carpentieri, Mario and Finocchio, Giovanni and Azzerboni, Bruno and Torres, Luis},
	month = sep,
	year = {2010},
	pages = {094434},
}

@article{taniguchi2016,
	title = {Magnetization switching by current and microwaves},
	volume = {93},
	url = {https://link.aps.org/doi/10.1103/PhysRevB.93.014430},
	doi = {10.1103/PhysRevB.93.014430},
	number = {1},
	urldate = {2025-08-19},
	journal = {Physical Review B},
	author = {Taniguchi, Tomohiro and Saida, Daisuke and Nakatani, Yoshinobu and Kubota, Hitoshi},
	month = jan,
	year = {2016},
	pages = {014430},
}

@article{florez2008,
	title = {Effects of radio-frequency current on spin-transfer-torque-induced dynamics},
	volume = {78},
	url = {https://link.aps.org/doi/10.1103/PhysRevB.78.184403},
	doi = {10.1103/PhysRevB.78.184403},
	number = {18},
	urldate = {2025-08-19},
	journal = {Physical Review B},
	author = {Florez, S. H. and Katine, J. A. and Carey, M. and Folks, L. and Ozatay, O. and Terris, B. D.},
	month = nov,
	year = {2008},
	pages = {184403},
}

@article{suto2015,
	title = {Microwave-assisted switching of a single perpendicular magnetic tunnel junction nanodot},
	volume = {8},
	issn = {1882-0786},
	url = {https://iopscience.iop.org/article/10.7567/APEX.8.023001/meta},
	doi = {10.7567/APEX.8.023001},
	language = {en},
	number = {2},
	urldate = {2025-08-19},
	journal = {Applied Physics Express},
	author = {Suto, Hirofumi and Nagasawa, Tazumi and Kudo, Kiwamu and Mizushima, Koichi and Sato, Rie},
	month = jan,
	year = {2015},
	pages = {023001},
}

@article{gapihan2012,
	title = {Heating asymmetry induced by tunneling current flow in magnetic tunnel junctions},
	volume = {100},
	issn = {0003-6951},
	url = {https://doi.org/10.1063/1.4719663},
	doi = {10.1063/1.4719663},
	number = {20},
	urldate = {2025-08-24},
	journal = {Applied Physics Letters},
	author = {Gapihan, E. and Hérault, J. and Sousa, R. C. and Dahmane, Y. and Dieny, B. and Vila, L. and Prejbeanu, I. L. and Ducruet, C. and Portemont, C. and Mackay, K. and Nozières, J. P.},
	month = may,
	year = {2012},
	pages = {202410},
}

@article{amara2012,
	title = {Modelling of time-dependent dielectric barrier breakdown mechanisms in {MgO}-based magnetic tunnel junctions},
	volume = {45},
	issn = {0022-3727},
	url = {https://dx.doi.org/10.1088/0022-3727/45/29/295002},
	doi = {10.1088/0022-3727/45/29/295002},
	language = {en},
	number = {29},
	urldate = {2025-08-24},
	journal = {Journal of Physics D: Applied Physics},
	author = {Amara-Dababi, S and Bea, H and Sousa, R and Mackay, K and Dieny, B},
	month = jul,
	year = {2012},
	pages = {295002},
}

@article{ciesielski2012,
	title = {The {Poincaré}-{Bendixson} {Theorem}: from {Poincaré} to the {XXIst} century},
	volume = {10},
	issn = {1644-3616},
	shorttitle = {The {Poincaré}-{Bendixson} {Theorem}},
	url = {https://doi.org/10.2478/s11533-012-0110-y},
	doi = {10.2478/s11533-012-0110-y},
	language = {en},
	number = {6},
	urldate = {2025-08-24},
	journal = {Central European Journal of Mathematics},
	author = {Ciesielski, Krzysztof},
	month = dec,
	year = {2012},
	pages = {2110--2128},
}

@article{hirayama2015,
	title = {In-plane anisotropy of a nano-scaled magnetic tunnel junction with perpendicular magnetic easy axis},
	volume = {54},
	issn = {1347-4065},
	url = {https://iopscience.iop.org/article/10.7567/JJAP.54.04DM03/meta},
	doi = {10.7567/JJAP.54.04DM03},
	language = {en},
	number = {4S},
	urldate = {2025-08-24},
	journal = {Japanese Journal of Applied Physics},
	author = {Hirayama, Eriko and Kanai, Shun and Sato, Koji and Yamanouchi, Michihiko and Sato, Hideo and Ikeda, Shoji and Matsukura, Fumihiro and Ohno, Hideo},
	month = feb,
	year = {2015},
	pages = {04DM03},
}

@article{mazraati2016,
	title = {Free- and reference-layer magnetization modes versus in-plane magnetic field in a magnetic tunnel junction with perpendicular magnetic easy axis},
	volume = {94},
	url = {https://link.aps.org/doi/10.1103/PhysRevB.94.104428},
	doi = {10.1103/PhysRevB.94.104428},
	number = {10},
	urldate = {2025-08-24},
	journal = {Physical Review B},
	author = {Mazraati, Hamid and Le, Tuan Q. and Awad, Ahmad A. and Chung, Sunjae and Hirayama, Eriko and Ikeda, Shoji and Matsukura, Fumihiro and Ohno, Hideo and Åkerman, Johan},
	month = sep,
	year = {2016},
	pages = {104428},
}

@article{serpico2015,
	title = {Heteroclinic tangle phenomena in nanomagnets subject to time-harmonic excitations},
	volume = {117},
	issn = {0021-8979},
	url = {https://doi.org/10.1063/1.4914530},
	doi = {10.1063/1.4914530},
	number = {17},
	urldate = {2025-08-24},
	journal = {Journal of Applied Physics},
	author = {Serpico, C. and Quercia, A. and Bertotti, G. and d'Aquino, M. and Mayergoyz, I. and Perna, S. and Ansalone, P.},
	month = mar,
	year = {2015},
	pages = {17B719},
}

@article{serpico2015_2,
	title = {Noise-induced bifurcations in magnetization dynamics of uniaxial nanomagnets},
	volume = {117},
	issn = {0021-8979},
	url = {https://doi.org/10.1063/1.4906961},
	doi = {10.1063/1.4906961},
	number = {17},
	urldate = {2025-08-24},
	journal = {Journal of Applied Physics},
	author = {Serpico, C. and Perna, S. and Bertotti, G. and d'Aquino, M. and Quercia, A. and Mayergoyz, I. D.},
	month = {02},
	year = {2015},
	pages = {17A709},
}

@article{suarez2017,
	title = {Chaotic dynamics of a magnetic particle at finite temperature},
	volume = {95},
	url = {https://link.aps.org/doi/10.1103/PhysRevB.95.014404},
	doi = {10.1103/PhysRevB.95.014404},
	number = {1},
	urldate = {2025-08-24},
	journal = {Physical Review B},
	author = {Suarez, O. J. and Laroze, D. and Martínez-Mardones, J. and Altbir, D. and Chubykalo-Fesenko, O.},
	month = jan,
	year = {2017},
	pages = {014404},
}

@article{ferona2017,
	title = {Nonlinear and chaotic magnetization dynamics near bifurcations of the {Landau}-{Lifshitz}-{Gilbert} equation},
	volume = {95},
	url = {https://link.aps.org/doi/10.1103/PhysRevB.95.104421},
	doi = {10.1103/PhysRevB.95.104421},
	number = {10},
	urldate = {2025-08-24},
	journal = {Physical Review B},
	author = {Ferona, Aaron M. and Camley, Robert E.},
	month = mar,
	year = {2017},
	pages = {104421},
}

@article{thirion2003,
	title = {Switching of magnetization by nonlinear resonance studied in single nanoparticles},
	volume = {2},
	copyright = {2003 Springer Nature Limited},
	issn = {1476-4660},
	url = {https://www.nature.com/articles/nmat946},
	doi = {10.1038/nmat946},
	language = {en},
	number = {8},
	urldate = {2025-08-24},
	journal = {Nature Materials},
	author = {Thirion, Christophe and Wernsdorfer, Wolfgang and Mailly, Dominique},
	month = aug,
	year = {2003},
	pages = {524--527},
}

@article{woltersdorf2007,
	title = {Microwave {Assisted} {Switching} of {Single} {Domain} \$\{{\textbackslash}mathrm\{{Ni}\}\}\_\{80\}\{{\textbackslash}mathrm\{{Fe}\}\}\_\{20\}\$ {Elements}},
	volume = {99},
	url = {https://link.aps.org/doi/10.1103/PhysRevLett.99.227207},
	doi = {10.1103/PhysRevLett.99.227207},
	number = {22},
	urldate = {2025-08-24},
	journal = {Physical Review Letters},
	author = {Woltersdorf, Georg and Back, Christian H.},
	month = nov,
	year = {2007},
	pages = {227207},
}

@article{kwiatkowski2021,
	title = {Optimal {Control} of {Magnetization} {Reversal} in a {Monodomain} {Particle} by {Means} of {Applied} {Magnetic} {Field}},
	volume = {126},
	url = {https://link.aps.org/doi/10.1103/PhysRevLett.126.177206},
	doi = {10.1103/PhysRevLett.126.177206},
	number = {17},
	urldate = {2025-08-24},
	journal = {Physical Review Letters},
	author = {Kwiatkowski, Grzegorz J. and Badarneh, Mohammad H. A. and Berkov, Dmitry V. and Bessarab, Pavel F.},
	month = apr,
	year = {2021},
	pages = {177206},
}

@article{montoya2020,
	title = {Immunity of nanoscale magnetic tunnel junctions with perpendicular magnetic anisotropy to ionizing radiation},
	volume = {10},
	copyright = {2020 The Author(s)},
	issn = {2045-2322},
	url = {https://www.nature.com/articles/s41598-020-67257-2},
	doi = {10.1038/s41598-020-67257-2},
	language = {en},
	number = {1},
	urldate = {2021-05-18},
	journal = {Scientific Reports},
	author = {Montoya, Eric Arturo and Chen, Jen-Ru and Ngelale, Randy and Lee, Han Kyu and Tseng, Hsin-Wei and Wan, Lei and Yang, En and Braganca, Patrick and Boyraz, Ozdal and Bagherzadeh, Nader and Nilsson, Mikael and Krivorotov, Ilya N.},
	month = jun,
	year = {2020},
	pages = {10220},
}

@article{safranski2016,
	title = {Material parameters of perpendicularly magnetized tunnel junctions from spin torque ferromagnetic resonance techniques},
	volume = {109},
	issn = {0003-6951},
	url = {https://aip.scitation.org/doi/abs/10.1063/1.4963354},
	doi = {10.1063/1.4963354},
	number = {13},
	urldate = {2019-07-22},
	journal = {Applied Physics Letters},
	author = {Safranski, Christopher J. and Chen, Yu-Jin and Krivorotov, Ilya N. and Sun, Jonathan Z.},
	month = sep,
	year = {2016},
	pages = {132408},
}

@article{finocchio2006,
	title = {Magnetization dynamics driven by the combined action of ac magnetic field and dc spin-polarized current},
	volume = {99},
	issn = {0021-8979},
	doi = {10.1063/1.2165136},
	journal = {J. Appl. Phys.},
	author = {Finocchio, G. and Krivorotov, I. and Carpentieri, M. and Consolo, G. and Azzerboni, B. and Torres, L. and Martinez, E. and Lopez-Diaz, L.},
	year = {2006},
	pages = {08G507},
}

@article{barsukov2014,
	title = {Field-dependent perpendicular magnetic anisotropy in {CoFeB} thin films},
	volume = {105},
	issn = {0003-6951, 1077-3118},
	url = {http://scitation.aip.org/content/aip/journal/apl/105/15/10.1063/1.4897939},
	doi = {10.1063/1.4897939},
	number = {15},
	urldate = {2015-10-18},
	journal = {Applied Physics Letters},
	author = {Barsukov, I. and Fu, Yu and Gonçalves, A. M. and Spasova, M. and Farle, M. and Sampaio, L. C. and Arias, R. E. and Krivorotov, I. N.},
	month = oct,
	year = {2014},
	pages = {152403},
}

@article{barsukov2019,
	title = {Giant nonlinear damping in nanoscale ferromagnets},
	volume = {5},
	copyright = {Copyright © 2019 The Authors, some rights reserved; exclusive licensee American Association for the Advancement of Science. No claim to original U.S. Government Works. Distributed under a Creative Commons Attribution NonCommercial License 4.0 (CC BY-NC).. This is an open-access article distributed under the terms of the Creative Commons Attribution-NonCommercial license, which permits use, distribution, and reproduction in any medium, so long as the resultant use is not for commercial advantage and provided the original work is properly cited.},
	issn = {2375-2548},
	url = {https://advances.sciencemag.org/content/5/10/eaav6943},
	doi = {10.1126/sciadv.aav6943},
	language = {en},
	number = {10},
	urldate = {2021-05-18},
	journal = {Science Advances},
	author = {Barsukov, I. and Lee, H. K. and Jara, A. A. and Chen, Y.-J. and Gonçalves, A. M. and Sha, C. and Katine, J. A. and Arias, R. E. and Ivanov, B. A. and Krivorotov, I. N.},
	month = oct,
	year = {2019},
	pages = {eaav6943},
}

@article{cheng2013,
	title = {Nonlinear ferromagnetic resonance induced by spin torque in nanoscale magnetic tunnel junctions},
	volume = {103},
	issn = {0003-6951},
	url = {https://aip.scitation.org/doi/full/10.1063/1.4819179},
	doi = {10.1063/1.4819179},
	number = {8},
	urldate = {2021-11-25},
	journal = {Applied Physics Letters},
	author = {Cheng, X. and Katine, J. A. and Rowlands, G. E. and Krivorotov, I. N.},
	month = aug,
	year = {2013},
	pages = {082402},
}

@article{chen2017,
	title = {Parametric {Resonance} of {Magnetization} {Excited} by {Electric} {Field}},
	volume = {17},
	issn = {1530-6984},
	url = {https://doi.org/10.1021/acs.nanolett.6b04725},
	doi = {10.1021/acs.nanolett.6b04725},
	number = {1},
	urldate = {2019-04-13},
	journal = {Nano Letters},
	author = {Chen, Yu-Jin and Lee, Han Kyu and Verba, Roman and Katine, Jordan A. and Barsukov, Igor and Tiberkevich, Vasil and Xiao, John Q. and Slavin, Andrei N. and Krivorotov, Ilya N.},
	month = jan,
	year = {2017},
	pages = {572--577},
}

@article{cubukcu2018,
	title = {Ultra-{Fast} {Perpendicular} {Spin}–{Orbit} {Torque} {MRAM}},
	volume = {54},
	issn = {1941-0069},
	doi = {10.1109/TMAG.2017.2772185},
	number = {4},
	journal = {IEEE Transactions on Magnetics},
	author = {Cubukcu, Murat and Boulle, Olivier and Mikuszeit, Nikolaï and Hamelin, Claire and Brächer, Thomas and Lamard, Nathalie and Cyrille, Marie-Claire and Buda-Prejbeanu, Liliana and Garello, Kevin and Miron, Ioan Mihai and Klein, O. and de Loubens, G. and Naletov, V. V. and Langer, Juergen and Ocker, Berthold and Gambardella, Pietro and Gaudin, Gilles},
	month = apr,
	year = {2018},
	pages = {1--4},
}

@article{marins_de_castro2012,
	title = {Precessional spin-transfer switching in a magnetic tunnel junction with a synthetic antiferromagnetic perpendicular polarizer},
	volume = {111},
	issn = {0021-8979},
	url = {https://doi.org/10.1063/1.3676610},
	doi = {10.1063/1.3676610},
	number = {7},
	urldate = {2025-08-24},
	journal = {Journal of Applied Physics},
	author = {Marins de Castro, M. and Sousa, R. C. and Bandiera, S. and Ducruet, C. and Chavent, A. and Auffret, S. and Papusoi, C. and Prejbeanu, I. L. and Portemont, C. and Vila, L. and Ebels, U. and Rodmacq, B. and Dieny, B.},
	month = mar,
	year = {2012},
	pages = {07C912},
}

@article{yamamoto2022,
	title = {Developments in voltage-controlled subnanosecond magnetization switching},
	volume = {560},
	issn = {0304-8853},
	url = {https://www.sciencedirect.com/science/article/pii/S0304885322005509},
	doi = {10.1016/j.jmmm.2022.169637},
	urldate = {2025-08-24},
	journal = {Journal of Magnetism and Magnetic Materials},
	author = {Yamamoto, Tatsuya and Matsumoto, Rie and Nozaki, Takayuki and Imamura, Hiroshi and Yuasa, Shinji},
	month = oct,
	year = {2022},
	pages = {169637},
}

@article{khanal2021,
	title = {Perpendicular magnetic tunnel junctions with multi-interface free layer},
	volume = {119},
	issn = {0003-6951},
	url = {https://doi.org/10.1063/5.0066782},
	doi = {10.1063/5.0066782},
	number = {24},
	urldate = {2025-08-25},
	journal = {Applied Physics Letters},
	author = {Khanal, Pravin and Zhou, Bowei and Andrade, Magda and Dang, Yanliu and Davydov, Albert and Habiboglu, Ali and Saidian, Jonah and Laurie, Adam and Wang, Jian-Ping and Gopman, Daniel B and Wang, Weigang},
	month = dec,
	year = {2021},
	pages = {242404},
}

@article{rehm2021,
	title = {Thermal {Effects} in {Spin}-{Torque} {Switching} of {Perpendicular} {Magnetic} {Tunnel} {Junctions} at {Cryogenic} {Temperatures}},
	volume = {15},
	url = {https://link.aps.org/doi/10.1103/PhysRevApplied.15.034088},
	doi = {10.1103/PhysRevApplied.15.034088},
	number = {3},
	urldate = {2025-08-25},
	journal = {Physical Review Applied},
	author = {Rehm, L. and Wolf, G. and Kardasz, B. and Cogulu, E. and Chen, Y. and Pinarbasi, M. and Kent, A.D.},
	month = mar,
	year = {2021},
	pages = {034088},
}

@article{mckinnon2022,
	title = {Thermally robust synthetic antiferromagnetic fixed layers containing {FeCoB} for use in {STT}-{MRAM} devices},
	volume = {546},
	issn = {0304-8853},
	url = {https://www.sciencedirect.com/science/article/pii/S0304885321008829},
	doi = {10.1016/j.jmmm.2021.168646},
	urldate = {2025-08-25},
	journal = {Journal of Magnetism and Magnetic Materials},
	author = {McKinnon, Tommy and Heinrich, Brett and Girt, Erol},
	month = mar,
	year = {2022},
	pages = {168646},
}

@article{beleggia_equivalent_2006,
	title = {The equivalent ellipsoid of a magnetized body},
	volume = {39},
	issn = {0022-3727, 1361-6463},
	url = {https://iopscience.iop.org/article/10.1088/0022-3727/39/5/001},
	doi = {10.1088/0022-3727/39/5/001},
	language = {en},
	number = {5},
	urldate = {2025-09-12},
	journal = {Journal of Physics D: Applied Physics},
	author = {Beleggia, M and Graef, M De and Millev, Y T},
	month = mar,
	year = {2006},
	pages = {891--899},
}

@article{kittel_theory_1948,
	title = {On the {Theory} of {Ferromagnetic} {Resonance} {Absorption}},
	volume = {73},
	copyright = {http://link.aps.org/licenses/aps-default-license},
	issn = {0031-899X},
	url = {https://link.aps.org/doi/10.1103/PhysRev.73.155},
	doi = {10.1103/PhysRev.73.155},
	language = {en},
	number = {2},
	urldate = {2025-09-12},
	journal = {Physical Review},
	author = {Kittel, Charles},
	month = jan,
	year = {1948},
	pages = {155--161},
}

@article{heinrich_ultrathin_1993,
	title = {Ultrathin metallic magnetic films: magnetic anisotropies and exchange interactions},
	volume = {42},
	issn = {0001-8732, 1460-6976},
	shorttitle = {Ultrathin metallic magnetic films},
	url = {http://www.tandfonline.com/doi/abs/10.1080/00018739300101524},
	doi = {10.1080/00018739300101524},
	language = {en},
	number = {5},
	urldate = {2025-09-13},
	journal = {Advances in Physics},
	author = {Heinrich, B. and Cochran, J.F.},
	month = oct,
	year = {1993},
	pages = {523--639},
}

@incollection{pozar_transients_2012,
	address = {Hoboken, NJ},
	edition = {4th ed},
	title = {Transients on {Transmission} {Lines}},
	isbn = {978-0-470-63155-3},
	booktitle = {Microwave engineering},
	publisher = {Wiley},
	author = {Pozar, David M.},
	year = {2012},
	pages = {85--90},
}

@article{sun_spin-torque_2013,
	title = {Spin-torque switching efficiency in {CoFeB}-{MgO} based tunnel junctions},
	volume = {88},
	copyright = {http://link.aps.org/licenses/aps-default-license},
	issn = {1098-0121, 1550-235X},
	url = {https://link.aps.org/doi/10.1103/PhysRevB.88.104426},
	doi = {10.1103/PhysRevB.88.104426},
	language = {en},
	number = {10},
	urldate = {2026-02-03},
	journal = {Physical Review B},
	author = {Sun, J. Z. and Brown, S. L. and Chen, W. and Delenia, E. A. and Gaidis, M. C. and Harms, J. and Hu, G. and Jiang, Xin and Kilaru, R. and Kula, W. and Lauer, G. and Liu, L. Q. and Murthy, S. and Nowak, J. and O’Sullivan, E. J. and Parkin, S. S. P. and Robertazzi, R. P. and Rice, P. M. and Sandhu, G. and Topuria, T. and Worledge, D. C.},
	month = sep,
	year = {2013},
	pages = {104426},
}

@article{wang_different_2014,
	title = {Different dielectric breakdown mechanisms for {RF}-{MgO} and naturally oxidized {MgO}},
	volume = {7},
	copyright = {http://iopscience.iop.org/info/page/text-and-data-mining},
	issn = {1882-0778, 1882-0786},
	url = {https://iopscience.iop.org/article/10.7567/APEX.7.083002},
	doi = {10.7567/APEX.7.083002},
	language = {en},
	number = {8},
	urldate = {2026-01-30},
	journal = {Applied Physics Express},
	author = {Wang, Xiaobin and Wang, Zihui and Hao, Xiaojie and Zhou, Yuchen and Zhang, Jing and Gan, Huadong and Jung, Dong Ha and Satoh, Kimihiro and Yen, Bing and Malmhall, Roger and Huai, Yiming},
	month = aug,
	year = {2014},
	pages = {083002},
}

@article{chavent_2016,
	title = {Steady {State} and {Dynamics} of {Joule} {Heating} in {Magnetic} {Tunnel} {Junctions} {Observed} via the {Temperature} {Dependence} of {RKKY} {Coupling}},
	volume = {6},
	url = {https://link.aps.org/doi/10.1103/PhysRevApplied.6.034003},
	doi = {10.1103/PhysRevApplied.6.034003},
	number = {3},
	urldate = {2026-02-23},
	journal = {Physical Review Applied},
	publisher = {American Physical Society},
	author = {Chavent, A. and Ducruet, C. and Portemont, C. and Vila, L. and Alvarez-Hérault, J. and Sousa, R. and Prejbeanu, I. L. and Dieny, B.},
	month = sep,
	year = {2016},
	pages = {034003},
}

@article{sullivan_1968,
	title = {Steady-{State}, ac-{Temperature} {Calorimetry}},
	volume = {173},
	url = {https://link.aps.org/doi/10.1103/PhysRev.173.679},
	doi = {10.1103/PhysRev.173.679},
	number = {3},
	urldate = {2026-02-23},
	journal = {Physical Review},
	publisher = {American Physical Society},
	author = {Sullivan, Paul F. and Seidel, G.},
	month = sep,
	year = {1968},
	pages = {679--685},
}

@book{Rogalski2010,
  author    = {Antoni Rogalski},
  title     = {Infrared Detectors},
  edition   = {2},
  publisher = {CRC Press},
  address   = {Boca Raton, FL},
  year      = {2010},
  isbn      = {978-1420076714}
}

@article{Beleggia2005,
title = "Demagnetization factors for elliptic cylinders",
author = "M Beleggia and \{De Graef\}, M and Millev, \{Y T\} and Goode, \{D A\} and G Rowlands",
year = "2005",
doi = "10.1088/0022-3727/38/18/001",
language = "English",
volume = "38",
pages = "3333--3342",
journal = "Journal of Physics D: Applied Physics",
issn = "0022-3727",
publisher = "IOP Publishing",
number = "18",
}
\end{document}


\title{\myfont Supplementary Material\\ Radio-frequency assisted switching in perpendicular magnetic tunnel junctions}

\author{\myfont Mark Hayward$^{1,}$}
\author{\myfont Salvatore Perna$^{2,}$}
\author{\myfont Massimiliano d'Aquino$^{2,}$}
\author{\myfont Claudio Serpico$^{2,}$}
\author{\myfont Wonjoon Jung$^{3,}$}
\author{\myfont Chunhui Dai$^{3}$}
\author{\myfont Patrick M. Braganca$^{3}$}
\author{\myfont Ilya N. Krivorotov$^{1,}$}

\affiliation{\myfont $^1$ Department of Physics and Astronomy, University of California, Irvine, California 92697, USA}
\affiliation{\myfont $^2$ Department of Electrical Engineering and Information Technology, University of Naples Federico II, Naples, Italy.}
\affiliation{\myfont $^3$ Western Digital Technologies, San Jose, California, 95119, USA}


\keywords{}
\maketitle

\section{RF-assisted switching Measurement}\label{sec:switching}
\subsection{Measurement Procedure}
\begin{figure*}[b]
\centering
 \includegraphics[width= 1.0\textwidth]{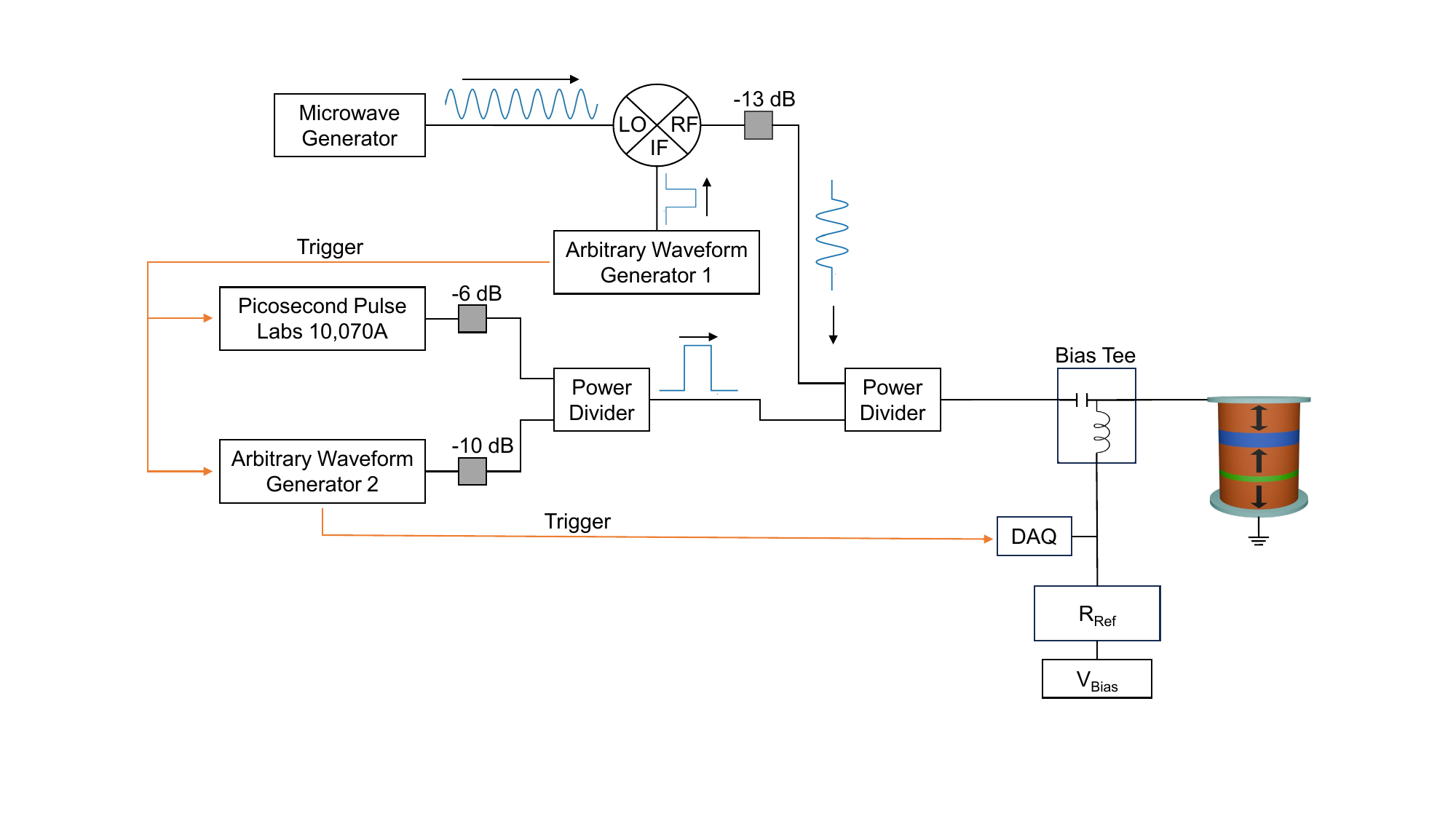}%
 \caption{\textbf{Circuit diagram used to create combined RF-DC pulse waveforms.} 
\label{fig:circuit}}%
\end{figure*}

We use the circuit in \cref{fig:circuit} to generate waveforms for current-induced switching of a magnetic tunnel junction (MTJ) free layer (FL), which consists of a combination of direct current (DC) and radio frequency (RF) pulses.  
Switching is detected by a change in MTJ resistance, monitored through a voltage divider formed with a fixed resistor $R_{Ref}$ and the MTJ.  
We bias the voltage divider with a low-level DC voltage $V_{\mathrm{Bias}}$ and use a 16-bit 2 MHz data acquisition board (DAQ) to measure time-dependent voltage across the MTJ.  
To switch the FL, we first apply an RF voltage pulse to the MTJ followed by a DC voltage pulse from a Picosecond Pulse Labs 10,070A pulse generator.  
Following the RF+DC write sequence, we use the DAQ to measure the resultant MTJ resistance, which registers either switching or non-switching. 
Then Arbitrary Waveform Generator 2 sends a lower-amplitude, longer-duration DC reset pulse of opposite polarity to the DC write pulse to return the MTJ to its initial state, and the DAQ confirms a successful reset.
The RF pulse is generated by using a broadband microwave mixer connected to a microwave generator, which controls the RF frequency, and to Arbitrary Waveform Generator 1, which controls the output RF pulse duration.  
The amplitude of the RF pulse is controlled by both generators. The RF, DC write, and DC reset pulses are combined into one waveform via two power dividers.  
Attenuators placed throughout the circuit suppress the background signal from the mixer and protect the Picosecond Pulse Labs 10,070A output port from overloading. 

The Picosecond Pulse Labs 10,070A and Arbitrary Waveform Generator 2 are triggered by the SYNC output of Arbitrary Waveform Generator 1.  
The DAQ is triggered by the SYNC output of Arbitrary Waveform Generator 2. 
The microwave generator operates in continuous wave mode, resulting in a random phase $\phi$ of the RF pulse.  
A delay $\tau$ between the RF and DC pulses is introduced and controlled by a combination of cable length and the programmable internal delay of the Picosecond Pulse Labs 10,070A.  
The switching probability $\mathcal{P}$ is determined as a ratio of the number of successful switching events to the total number of switching attempts. 
Each measurement is repeated $10^5$ times to reliably measure $\mathcal{P}$.  

\subsection{Waveform Characterization}
\indent The pulse amplitudes of the RF+DC waveforms are characterized by replacing the MTJ in \cref{fig:circuit} with a high-bandwidth, real-time oscilloscope.  However, because the oscilloscope has a $50\,\Omega$ input impedance, the scope does not account for the large impedance mismatch between the high impedance MTJ and the lower impedance circuit. This mismatch causes the pulse amplitudes delivered to the MTJ to nearly double in amplitude compared to the amplitude measured by the oscilloscope. The impedance mismatch and the subsequent increase in pulse amplitude must be taken into account during circuit calibration \cite{pozar_transients_2012}.
The root mean square (RMS) voltage of the RF pulse $V_{\mathrm{RF}}$ delivered to the high-impedance MTJ is given by \cref{eq:Vrms}, where $V^{\mathrm{scope}}_{\mathrm{PP}}$ is the peak-to-peak voltage measured by the 50\,$\Omega$ oscilloscope.
\begin{equation}\label{eq:Vrms}
    V_{\mathrm{RF}}=\frac{2\cdot V^{\mathrm{scope}}_{\mathrm{PP}}}{2\sqrt{2}} = \frac{V^{\mathrm{scope}}_{\mathrm{PP}}}{\sqrt{2}}
\end{equation} 
Similarly, \cref{eq:Vdc} gives the amplitude of the DC write and reset pulse dissipated by the high impedance MTJ. $V^{\mathrm{scope}}_{\mathrm{DC}}$ is the amplitude of the DC write pulse measured by the oscilloscope.
\begin{equation}\label{eq:Vdc}
    V_{\mathrm{DC}}=2\cdot V^{\mathrm{scope}}_{\mathrm{DC}}
\end{equation}
\Cref{eq:Powercalc} defines RF power as the power dissipated by the sample (in dBm).  
\begin{equation}\label{eq:Powercalc}
    P_{\text{RF}}=10 \cdot \log_{10} \left( \frac{V_{\text{RF}}^2}{R_{\text{P}}}\cdot \frac{1}{1\, \text{mW}} \right)
\end{equation}  
Here, $V_{\text{RF}}$ is defined by \cref{eq:Vrms}, and $R_P$ is the resistance of the MTJ in the parallel resistance state.   All pulse amplitudes reported in this work are the amplitudes dissipated by the MTJ as described by \cref{eq:Vrms} and \cref{eq:Vdc}.  
{\cref{tab2} lists the measured values of $V_{\mathrm{RF}}$ for the frequencies $f_\mathrm{RF}$ used in Fig. 5 of the main text, ensuring the power dissipated by the MTJ $P_{\text{RF}}$ is uniform for all RF frequencies used.

\begin{table}[ht]
    \caption{Characterization of the RF voltage for various frequencies.
    \\}\label{tab2}
    \centering
    {
    \begin{tabular}{ c c }
        \hline
        $f_\mathrm{RF}$ (GHz) & $V_{\mathrm{RF}}$ (V) \\
        \hline
        $0.1$ & $0.180$ \\
        $0.5$ & $0.185$ \\
        $1$ & $0.181$ \\
        $4$ & $0.184$ \\
        \hline
    \end{tabular}}
\end{table}
\subsection{Energy calculations}\label{sec:energy_calc}

To compare energy efficiency of different voltage waveforms used for STT-MRAM switching, we compute the total energy dissipated by the MTJ due to the RF and DC voltage pulses $E_{\mathrm{tot}}$:
\begin{equation}\label{eq:energy_efficiency}
    E_{\mathrm{tot}}=\frac{V^2_\mathrm{RF}}{R_\mathrm{P}}t_\mathrm{RF}+\frac{V^2_\mathrm{DC}}{R_\mathrm{P}}t_\mathrm{DC}
\end{equation}
where $V_{\mathrm{RF}}$ is determined by \cref{eq:Vrms}, $V_{\mathrm{DC}}$ is determined by \cref{eq:Vdc}, $t_\mathrm{RF}$ is the RF pulse duration, and $t_\mathrm{DC}$ is the DC write pulse duration.
This expression is valid for both positive and negative delays $\tau$ when $ E_{\mathrm{tot}}$ is an ensemble average over random phases of the RF pulse, which is the case for our measurements.}
\begin{figure*}[ht]
\centering
 \includegraphics[width= 1.0\textwidth]{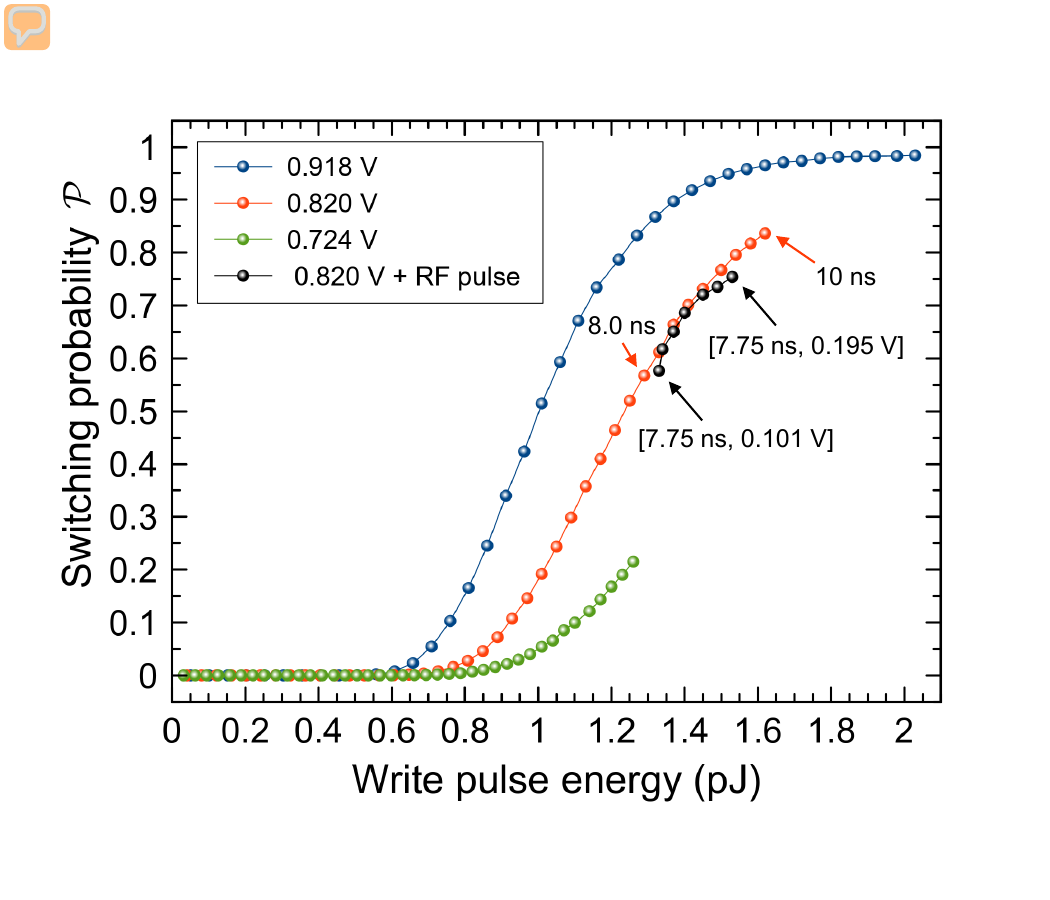}%
 \caption{\textbf{$\mathcal{P}$ as a function of total waveform energy $E_{\mathrm{tot}}$ for three DC pulse amplitudes and a combined RF+DC pulse.} Here $\tau = -7$\,ns, $D = 45$\,nm, and $f_{\mathrm{RF}}=0.1$\,GHz.  
 The red arrows label values of $t_{\mathrm{DC}}$ at a fixed $V_{\mathrm{DC}}$ for the DC-only waveform, while the black arrows label [$t_{\mathrm{DC}}$,$V_{\mathrm{RF}}$] for the RF+DC waveform.}
\label{fig:energy}%
\end{figure*}
\cref{fig:energy} shows switching probability $\mathcal{P}$ as a function of write pulse energy for a $D=45$\,nm device, comparing three DC-only write pulse amplitudes (blue, red, green) with an RF+DC write waveform (black). 
In the RF+DC configuration, the fixed values of $V_{\mathrm{DC}} = 0.820$\,V and $t_{\mathrm{DC}} = 7.75$\,ns are chosen such that $\mathcal{P} =0.5$ in the absence of an RF pulse. Here, the delay $\tau = -7$\,ns, and $f_{\mathrm{RF}}=0.1$\,GHz.  
The red arrows label two values of $t_{\mathrm{DC}}$ at a fixed $V_{\mathrm{DC}}$ for the DC-only waveform, while the black arrows label two values of $V_{\mathrm{RF}}$ for the RF+DC waveform.  
As shown in \cref{fig:energy}, the RF+DC write waveform attains nearly the same total write energy as the DC-only pulse with the same amplitude $V_{\mathrm{DC}}$  but with a reduced $t_{\mathrm{DC}}$.  

\subsection{Measurement Uncertainty Calculations}\label{sec:energy_calc}
To establish confidence in the data presented in Fig.\,5 of the main text, we analyze the statistical uncertainty associated with each measurement. 
Because each switching attempt results in one of two outcomes -- successful or unsuccessful switching -- the data are described by a binomial distribution.  
We assume independence between switching events.  
Each probability shown in Fig.\,5 is measured from $10^6$ switching attempts.  
The change in switching probability $\Delta \mathcal{P}$ is defined as:
\begin{equation}\label{eq:delP}
    \Delta \mathcal{P} = \mathcal{P_{\mathrm{RF}}} - \mathcal{\overline{P}}_{\mathrm{DC}}
\end{equation}
where $\mathcal{P_{\mathrm{RF}}}$ is the switching probability for the RF+DC write pulse, and $\mathcal{\overline{P}}_{\mathrm{DC}}$ is the mean switching probability for the DC only write pulse, averaged over 17 independent measurements (each determined by $10^6$ switching events). 
The error bars in Fig. 5 of the main text represent $\pm$ the uncertainty $\sigma_{\Delta\mathcal{P}}$ in $\Delta\mathcal{P}$ and is obtained by propagating the uncertainties in $\mathcal{P}_{\mathrm{RF}}$ and $\mathcal{\overline{P}}_{\mathrm{DC}}$:
\begin{equation}\label{eq:errordelp}
\sigma_{\Delta\mathcal{P}}=\sqrt{\sigma_{\mathcal{P_{\mathrm{RF}}}}^2+\sigma_{\mathcal{\overline{P}}_{\mathrm{DC}}}^2}
\end{equation}
$\sigma_{\mathcal{P_{\mathrm{RF}}}}$ is the uncertainty in the RF+DC write pulse measurements defined by the binomial distribution: 
\begin{equation}
    \sigma_{\mathcal{P_{\mathrm{RF}}}}=\sqrt{\frac{\mathcal{P_{\mathrm{RF}}}(1-\mathcal{P_{\mathrm{RF}}})}{N}}
\end{equation}
with $N = 10^6$ switching events, and $\sigma_{\mathcal{\overline{P}_{\mathrm{DC}}}}$ is standard error of the mean for the DC pulse measurements:
\begin{equation}
    \sigma_{{\mathcal{\overline{P}}_{\mathrm{DC}}}}=\sqrt{\frac{s^2}{n}}
\end{equation}
where $n = 17$ is the number of measurements without the RF pulse, and $s^2$ is the sample variance:
\begin{equation}\label{samplevar}
    s^2=\frac{1}{n-1}\sum\limits_{i=1}^{n}(\mathcal{P_{\mathrm{DC,i}}}-\mathcal{\overline{P}}_{\mathrm{DC}})^2
\end{equation}

\section{Spin torque ferromagnetic resonance}\label{sec:stfmr}
\begin{figure*}[t]
\centering
 \includegraphics[width= 1.0\textwidth]{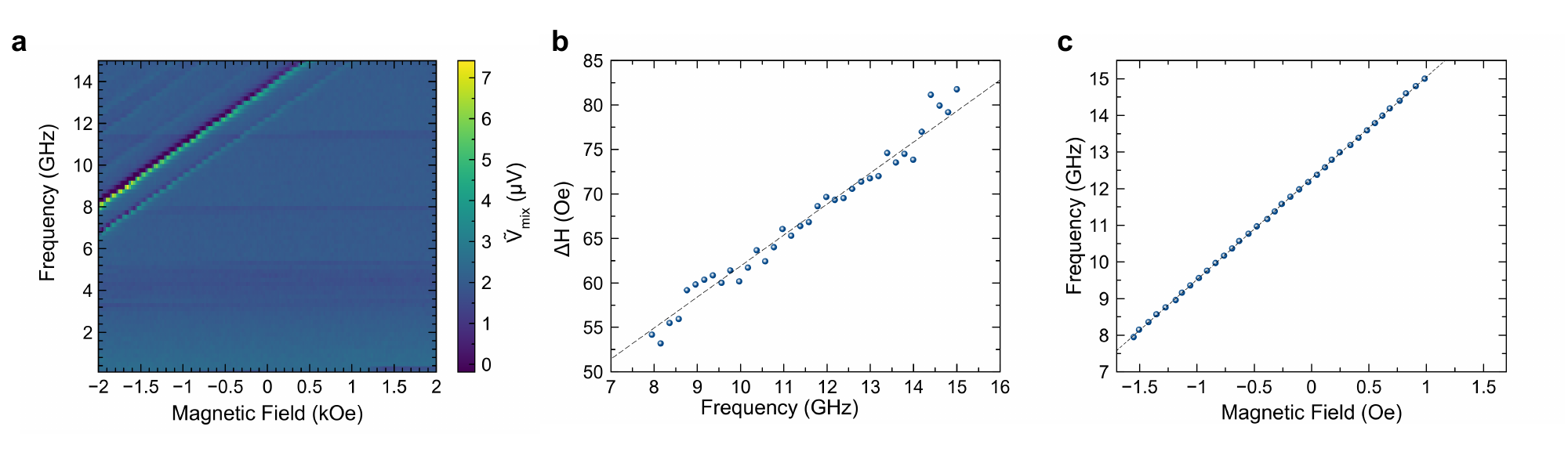}%
 \caption{\textbf{Spin torque ferromagnetic resonance measurements for a $D=85$\,nm device.} \textbf{a} $\tilde{V}_{\mathrm{mix}}$ as a function of frequency and magnetic field. 
 \textbf{b} Linewidth $\Delta H$ defined as the HWHM of the resonance peak as a function of frequency fitted to \cref{eq:linewidth}  \textbf{c} Frequency as a function of magnetic field where data points represent the resonance field for a given frequency fitted to \cref{eq:kittelequation}.
\label{fig:ST-FMR}}%
\end{figure*}

We use spin torque ferromagnetic resonance (ST-FMR) to measure the frequency and linewidth of the free layer (FL) spin wave modes as a function of out-of-plane magnetic field. 
From these data, we extract the FL Gilbert damping constant $\alpha$ and the uniaxial magnetocrystalline anisotropy constant $K_{AN}$ at room temperature \cite{gonccalves2013spin, safranski2016, barsukov2019}.  
In these measurements, a microwave voltage is applied to the MTJ via the RF port of a bias tee.  
The applied voltage applies an RF spin transfer torque that excites spin waves in the FL and leads to oscillations of the sample resistance at the drive frequency due to the tunneling magnetoresistance of the MTJ. 
These resistance oscillations mix with the RF current through the MTJ and generate a rectified DC voltage $V_{\mathrm{mix}}$, which is measured by a lock-in amplifier through the DC port of the bias tee. 
The magnetic field modulation technique is employed to improve the signal-to-noise ratio \cite{gonccalves2013spin}.  
The measured signal $\tilde{V}_{\mathrm{mix}}$ is proportional to the magnetic field derivative of the rectified DC voltage.
The linewidth $\Delta H$, defined as the half-width at half-maximum (HWHM) of the Lorentzian function, is obtained by fitting the resonance peaks to a sum of magnetic field derivatives of symmetric and anti-symmetric Lorentzian functions \cite{gonccalves2013spin}. 
\Cref{fig:ST-FMR}a shows ST-FMR spectra for a $D=85$\,nm device, while \cref{fig:ST-FMR}b shows $\Delta H$ as a function of frequency.  
The data in \cref{fig:ST-FMR}b are fit using \cref{eq:linewidth} to extract the Gilbert damping constant $\alpha$:
\begin{equation}\label{eq:linewidth}
    \Delta H(\omega)=\alpha \frac{\omega}{\gamma}+\Delta H(0)
\end{equation}
where $\omega=2\pi f$ is the microwave angular frequency, $\gamma$ is the gyromagnetic ratio, and $\Delta H(0)$ is the frequency-independent linewidth due to the presence of magnetic inhomogeneities \cite{heinrich_ultrathin_1993}.
\Cref{fig:ST-FMR}c depicts frequency of the quasi-uniform (lowest frequency) spin wave mode as a function of magnetic field.  
The data points represent the resonance field for a given frequency extracted from \cref{fig:ST-FMR}a.  
Using the easy axis approximation of the Kittel equation, we fit \cref{eq:kittelequation} to the data in \cref{fig:ST-FMR} to determine the effective magnetic anisotropy $H_{\mathrm{K}}$, from which the magnetocrystalline anisotropy constant $K_{\mathrm{AN}}$ is calculated:
\begin{equation}\label{eq:kittelequation}
    \omega=\gamma \left(H +H_{\mathrm{dip}}+H_{\mathrm{K}}\right)
\end{equation}
Here, $H_\mathrm{dip}$ is the center of the magnetic hysteresis loop, and the effective magnetic anisotropy is given by:
\begin{equation}\label{eq:anisotropy}
    H_{\mathrm{K}}=\frac{2K_{\mathrm{AN}}}{\mu_0 M_s}-( D_z-D_\mathrm{\perp}) M_s
\end{equation}
where $M_s$ is the saturation magnetization, $\mu_0$ is the magnetic permeability of free space, and $D_z$ and $D_\mathrm{\perp}$ are the demagnetizing factor along the symmetry axis and perpendicular direction, respectively. 
These factors satisfy the relation: $2D_\perp +D_z = 1$ \cite{kittel_theory_1948,beleggia_equivalent_2006,heinrich_ultrathin_1993}. Fitting \cref{eq:linewidth} and \cref{eq:kittelequation} to \cref{fig:ST-FMR}b,c respectively gives $\alpha = 0.0097$ and $K_{\mathrm{AN}} = 8.2\times 10^5$\,$\mathrm{J/m^3}$.

\bigskip

\section{Micromagnetic description of the switching dynamics}\label{sec:theory}
 \subsection{Micromagnetic model}
The magnetization dynamics are described by the stochastic Landau-Lifshitz-Gilbert-Slonczewski (LLGS) equation expressed in the following normalized form \cite{Serpico2008}:
\begin{equation}\label{eq:LLGS}
\frac{\partial \bm m}{\partial t} -\alpha\bm m\times\frac{\partial \bm m}{\partial t} = -\bm m\times\bm h_{\text{eff}} + \beta_\text{ST}\, \bm m\times(\bm m\times \bm p)-\nu\,\bm m\times\bm h_N\,,
\end{equation}
where $\bm m = \bm M/M_s$ is the magnetic unit vector given by the ratio of the magnetization vector and the saturation magnetization at a given value of the absolute temperature $T$, $t$ is the normalized time in units of $(\gamma M_s)^{-1}$ with $\gamma$ being the absolute value of the gyromagnetic ratio, $\alpha$ is the Gilbert damping, $\bm h_{\text{eff}} = \bm h_{\text{ex}}+\bm h_{\text{m}}+\bm h_{\text{an}}+\bm h_\text{a}$ is the normalized effective field defined as the sum of the exchange, magnetostatic, anisotropy, and applied fields respectively, $\beta_\text{ST}$ is the spin torque coefficient proportional to current, and $\bm p$ is the magnetization unit vector of the polarizer. 

The last torque term on the right-hand side is the stochastic torque acting on the magnetization field, where the coefficient $\nu$ satisfies the following fluctuation-dissipation relation: $\nu^2/(2\alpha) = k_BT/(\mu_0 M_s^2 V) = \tilde{T}$. 
Here, $k_B$ is the Boltzmann constant and $V$ is the volume of the free layer for macrospin simulations while it is the volume of the single cell for micromagnetic simulations. 
The thermal field $\bm h_N(t)$ is a Gaussian white noise whose components satisfy the following relations: $<h_{N,i}(t)\,h_{N,j}(t')> = \delta_{ij}\delta(t-t')$, where the angular brackets $<.>$ refer to the statistical average, $\delta_{ij}$ is the Kronecker delta and $\delta(t-t')$ is the Dirac delta function.

In the macrospin approximation, the magnetization is assumed to be spatially uniform. This produces zero exchange field. Moreover the magnetostatic field is assumed to be expressed by $\bm h_\text{m} = -\bm D\cdot \bm m$, where $\bm D$ is a diagonal tensor $(\mathrm{Tr}(\bm D) = 1)$ whose components are the demagnetizing factors. Under the macrospin approximation,  \cref{eq:LLGS} becomes an ordinary differential equation.
In micromagnetic simulations, the magnetization of the free layer is, in general, not uniform, the exchange field is not zero, unlike the macrospin approximation, and the magnetostatic field depends on the magnetization field distribution by a nonlocal operator. 
The time-dependent excitation is simulated by giving $\beta_{ST}$ the same time dependence of the voltage pulse waveform used in the experiments. The thermal field has been computed assuming the single replica is in contact with a thermal bath at room temperature $T=300\, K$.

\subsection{Setup for numerical computation of switching probability}
The switching probability is computed by integrating the stochastic magnetization dynamics and counting the switching events in an ensemble of replicas of the MTJ free layer (FL). We considered two ensembles: one consisting of $10^2$ replicas, where for each the stochastic magnetization dynamics is computed by a full micromagnetic solver. The second ensemble consists of $10^4$ replicas, where for each of them, the dynamics is computed by a macrospin solver. From the law of the large numbers, the error of the estimated probability is proportional to $1/\sqrt{N_\mathrm{R}}$, where $N_\mathrm{R}$ is the number of replicas. This means that such error is $\sim1\%$ for macrospin simulations, while it is  $\sim10\%$ for the micromagnetic ones.

All the simulations were performed by MaGICo (available at \url{http://wpage.unina.it/mdaquino/index_file/MaGICo.html}), which permits the integration of the stochastic LLGS equation in both full micromagnetic and macrospin approximation settings. The parameters used in the simulations are: FL thickness $= 2.21\, \mathrm{nm} $, saturation magnetization $ M_\mathrm{s} = 1039\,\mathrm{kA}/\mathrm{m}$, the anisotropy constant $K_\mathrm{AN} = 8.6\times10^5\, \mathrm{J}/\mathrm{m^3}$, the damping constant $\alpha = 0.0097$, and the exchange stiffness $A_\mathrm{ex} = 13\,\mathrm{pJ}/\mathrm{m}$. From them, the exchange length is computed to be $l_\mathrm{ex} = 4.38\,\mathrm{nm}$. The dimensions of the mesh used for micromagnetic analysis and the perpendicular demagnetizing factor $D_\perp$ for different FL diameters are reported in Table \ref{tab1}. 

\begin{table}[ht]
    \caption{Mesh dimensions and the demagnetizing factor $D_\perp$ for the several FL diameters.}\label{tab1}
    \centering
    {
    \begin{tabular}{ c c c c}
        \hline
        Diameter  &\hspace{0.5cm} Number of cells &\hspace{0.5cm} Cell volume & \hspace{0.5 cm} Demagnetizing \\
          (nm)    &\hspace{0.5cm} $\mathrm{N_x\times N_y\times N_z}$ & \hspace{0.5cm} $\mathrm{d_x\times d_y \times d_z}$ $(\mathrm{nm}^3)$ & \hspace{0.5cm} factor $D_\perp\, \mathrm{(ad.)}$\\
        \hline
        25 & $\mathrm{9\times 9\times 1}$ & $\mathrm{2.78\times 2.78\times 2.21}$   & 0.0948\\
        45 & $\mathrm{13\times 13\times 1}$ & $\mathrm{3.46\times 3.46\times 2.21}$ & 0.0620\\
        65 & $\mathrm{17\times 17\times 1}$ & $\mathrm{3.82\times 3.82\times 2.21}$ & 0.0474\\
        85 & $\mathrm{21\times 21\times 1}$ & $\mathrm{4.05\times 4.05\times 2.21}$ & 0.0387\\
        \hline
    \end{tabular}
    }
\end{table}
The demagnetizing factors that are used in the macrospin simulations have been estimated from the micromagnetic solver. They are in quantitative agreement with the values given in Ref. \cite{Beleggia2005}.

\subsection{Analytic theory of switching probability}
The discussion below establishes the theoretical basis for the analytical model and the assumptions (i)–(v) in the main text, presenting the explanations in order.

Numerical evidence supports the use of the macrospin model. More specifically, micromagnetic simulations agree quantitatively with ensemble simulations.\\
\indent Stochastic magnetization dynamics can be described in probabilistic terms by writing the Fokker-Planck (FP) equation corresponding to the stochastic LLGS equation. When cylindrical coordinates are used and the FP equation is averaged along the azimuthal coordinate, the following equation is obtained \cite{Brown1963}:
\begin{equation}
\frac{\partial p}{\partial t} = \frac{\partial}{\partial m_z}\left[\alpha(1-m_z^2)\left(\frac{\partial\Phi}{\partial m_z}p+\tilde{T}\frac{\partial p}{\partial m_z}\right)\right]\,,
\end{equation}
where $p(m_z,t)$ is the probability density function (PDF) described in Eq. (5) in the main text, and  $\Phi = k_\mathrm{eff}(1-m_z^2)/2 +(\beta_\mathrm{ST}/\alpha -h_\mathrm{az})m_z $ is an effective potential that includes the magnetic free energy and the spin-torque effect. The term $k_\mathrm{eff}$ is the normalized effective magnetic anisotropy constant of the FL, and it is given by $k_\mathrm{eff} = 2 K_\mathrm{AN}/(\mu_0 M_s^2) - (D_\mathrm{z}-D_\mathrm{\perp})$, where $K_\mathrm{AN}$ is the uniaxial magnetocrystalline anisotropy constant and $D_\mathrm{z}\,, D_\mathrm{\perp}$ are the demagnetizing factors along the symmetry axis and the perpendicular direction.\\
\indent Under the condition of stationary excitation, the distribution of the replicas will reach a steady state that is described by the following equilibrium PDF:
\begin{equation}\label{eq:eq_pdf}
    p_\mathrm{eq}(m_z) = \frac{1}{Z}\exp\left(-\frac{\Phi}{\tilde{T}}\right)\,,\,m_z\in[-1,1]\,,
\end{equation}
where $Z= \int_{-1}^1\exp\left(-\Phi/\tilde{T}\right) dm_z$ is the partition function, and its value is such that the probability verifies the normalization condition: $\int_{-1}^1 p(m_z)\,dm_z = 1$. When the replicas are not distributed according to $p_\mathrm{eq}(m_z)$, the PDF evolves according to a drift-diffusion process, where the ratio between $k_\mathrm{eff}/\tilde{T}$ gives us an indication about the character of the stochastic dynamics. In our case, $k_\mathrm{eff}/\tilde{T}\gg 1$.
This means that, when the external excitations are turned off, and all the replicas are in the well of one metastable state, relaxation towards the equilibrium PDF can be described by the Néel-Brown model in the high energy barrier case, with an initial distribution that is given by the following equation \cite{Brown1963}: 
\begin{equation}\label{eq:high_barr_pdf}
p_0(m_z) = \frac{1}{Z}\exp\left(\frac{k_\mathrm{eff}}{2\tilde{T}}m_z^2\right)\,,\,m_z\in[0,1]\,,
\end{equation}
where in this case $Z = \int_0^1\exp\left(\frac{k_\mathrm{eff}}{2\tilde{T}}m_z^2\right) dm_z$.\\ 
\indent Thermal relaxation to equilibrium occurs on a time scale that is much larger than the time window of the entire switching procedure ($\sim 100$\,ns).
When the voltage overcomes the threshold value and the magnetic state of a certain replica moves far from its deterministic equilibrium, the PDF dynamics are driven mostly by the drift process and therefore the magnetization dynamics can be considered deterministic.\\
\indent Ensemble simulations show that the switching probability does not change for $\tau \ge 0$. This result can be justified by investigating magnetization dynamics around the equilibrium. For $\tau \ge 0$, the RF voltage excites the system and the magnetic state moves far from equilibrium. However, the RF voltage is unable to switch the magnetic state on its own.  When the RF voltage is turned off, a free evolution occurs around one of the two equilibria.
When $\beta_\mathrm{ST}>0$, the time constant of the free evolution around the equilibrium can be estimated to be: $\tau_E = (2\gamma M_s \alpha\,\omega_K)^{-1}$, where $\omega_K = k_\mathrm{eff}+ h_{az}$ is the dimensionless Kittel frequency.
For $\beta_\mathrm{ST}\sim 0.5\, k_\mathrm{eff}$, the time constant of the free evolution around equilibrium is $\tau_E\sim 1$\,ns. However, experimentally we see influences of the RF voltage up to $\tau \approx 6$\,ns. 
This suggests that the mechanism behind the increase in switching probability for $\tau \ge 0$ originates elsewhere. 
If magnetization nonuniformities are considered, the free evolution time constant for a generic spin wave mode of wave number $k$ is $\tau_{E\mathrm{k}} = (2\gamma M_s  \alpha\,\omega_k)^{-1}\sim 1$\,ns, because $\omega_k\sim \omega_K$, where $\omega_k$ is the natural frequency of the $k$-th spin wave. 
This last consideration is also supported by numerical simulations (see main text), and it excludes the micromagnetic origin of the enhancement of switching probability for $\tau \ge 0$. \\
\indent For $\tau <0$, the RF voltage superimposes on the DC pulse, therefore $\beta_\text{ST} = \beta_\text{DC} + \beta_\text{RF}$. 
In a certain time window, the RF voltage can always be decomposed as $\beta_\mathrm{RF} =\overline{\beta}_\mathrm{RF} + \delta\beta_\mathrm{RF}(t')$, where $\overline{\beta}_\mathrm{RF}$ is the average value of the RF voltage on the DC pulse time window, and $t' = t-(t_\mathrm{RF}+\tau)$. 
If the RF power is small compared to the DC power, $|\beta_\mathrm{DC}+\overline{\beta}_\mathrm{RF}|\gg|\delta\beta_\mathrm{RF}(t')|$, the equation governing the dynamics of $m_z$ can be then written as:
\begin{equation}
\alpha k_\mathrm{eff}\,dt' = \frac{dm_z}{(1-m_z^2)(mz-H(t'))}\approx\frac{dm_z}{(1-m_z^2)(m_z-\overline{H})}\,,
\end{equation}
where $H(t') = \beta_\mathrm{ST}(t')/(\alpha k_\mathrm{eff})$ and $\overline{H} = (\beta_\mathrm{DC}+\overline{\beta}_\mathrm{RF})/(\alpha\,k_{\mathrm{eff}})$ is the effective DC voltage pulse amplitude normalized to the critical DC voltage for switching at zero temperature in dimensionless units.

The above arguments permit to write the PDF of the FL magnetization at an instant $t'$ as follows:
\begin{equation}\label{eq:prob_1}
    p(m_z,t') = \int_0^1 p_0(m_{z0})\,p(m_z,t'|m_{z0},0)\,dm_{z0}\,,
\end{equation}
where $m_{z0}$ is $m_z$ at $t' = 0$, which marks the onset of the $\overline{V}_\mathrm{ST}$ write pulse.

The magnetization switching dynamics is a damping-switching process in a uniaxial, bistable magnet between the two stable states $m_z = \pm 1$.
In this case the magnetization trajectory from  $m_{z0}$ to $m_z$ under the action of  $\overline{H} = \overline{V}_\mathrm{ST}/V_\mathrm{c0} = \overline{\beta}_\mathrm{ST}/(\alpha k_\mathrm{eff})$ can be expressed in closed form by the following \cite{dAquino2020}: $q(m_z,m_{z0},\overline{H}) = \alpha k_\mathrm{eff}t$. 
The quantity $\overline{H}$ is the effective DC voltage pulse amplitude normalized to $V_\mathrm{c0}$, which is the critical DC voltage for switching at $T = 0$. 
For the numerical integration of the magnetization dynamics, the dimensionless LLGS equation is considered \cite{Serpico2008}. 
Within it, the dimensionless voltage is $\beta_\mathrm{ST}$, while the dimensionless critical voltage at zero temperature is given by $\alpha k_\mathrm{eff}$.

Solving the relation defining $q$ for $m_z$, we obtain: $m_z(t) = \mu_z(m_{z0},t,\overline{H})$, which allows us to rewrite \cref{eq:prob_1} in the following form: 
\begin{equation}\label{eq:prob_2}
   p(m_z,t) = \frac{p_0(\mu_z^{-1}(m_z))}{\lambda_z(m_z)}\,,
\end{equation}
where $\lambda_z = d\mu_z/dm_{z0}$. 
We assume that the FL switches if $m_z$ changes sign compared to $m_{z0}$ after the write pulse. Starting with the initial distribution with $m_{z0} > 0$, the switching probability can be computed by integrating the PDF over the range $m_z\in[-1,0]$ that defines the well of the switched magnetic state $m_z = -1$.
Such a state cannot be reached by deterministic dynamics in a time window of finite duration. 
The closest state to $m_z = -1$ reachable in a finite time window is $m_z=\mu_z(0,t,\overline{H})$, which is the final state of a deterministic trajectory in the time interval $[0,t]$ starting from the initial state $m_z(0) = 0$.
With these considerations, the switching probability $\mathcal{P}$ at $t$ is given by:
\begin{widetext}
\begin{equation}\label{eq:prob_3}
\begin{aligned}
    \mathcal{P}(t') = \int_{\mu_z(0,t,\overline{H})}^0 p(m_z,t)\,dm_z =\int_0^{\mu_z^{-1}(0,t,\overline{H})}p_{0}(m_{z0})\,dm_{z0}\,
    = \frac{\mathrm{erfi}\left(\sqrt{\frac{k_\mathrm{eff}}{2\tilde{T}}}\mu_z^{-1}(0,t,\overline{H})\right)}{\mathrm{erfi}\left(\sqrt{\frac{k_\mathrm{eff}}{2\tilde{T}}}\right)}\,.
    \end{aligned}
\end{equation}
\end{widetext}
Taking into account that PDF depends on $\phi$ via $\overline{H}$, the switching probability for $\phi$ uniformly distributed over the experimentally relevant interval $[-\pi,\pi]$ is:
\begin{widetext}
\begin{equation}\label{eq:prob_4}
    \mathcal{P}(t') =\frac{\int_0^{2\pi}\mathrm{erfi}\left(\sqrt{\frac{k_\mathrm{eff}}{2\tilde{T}}}\mu_z^{-1}(0,t',\overline{H})\right)d\phi}{2\pi\,\mathrm{erfi}\left(\sqrt{\frac{k_\mathrm{eff}}{2\tilde{T}}}\right)}\,.
\end{equation}
\end{widetext}
This formula is of limited use to the case of low RF power, as can be seen by comparing Fig. 6d and Fig. 7c in the main text. Indeed, for high RF power the analytical estimation of the probability is not accurate anymore.

\begin{figure*}[ht]
\centering
 \includegraphics[width= 1.0\textwidth]{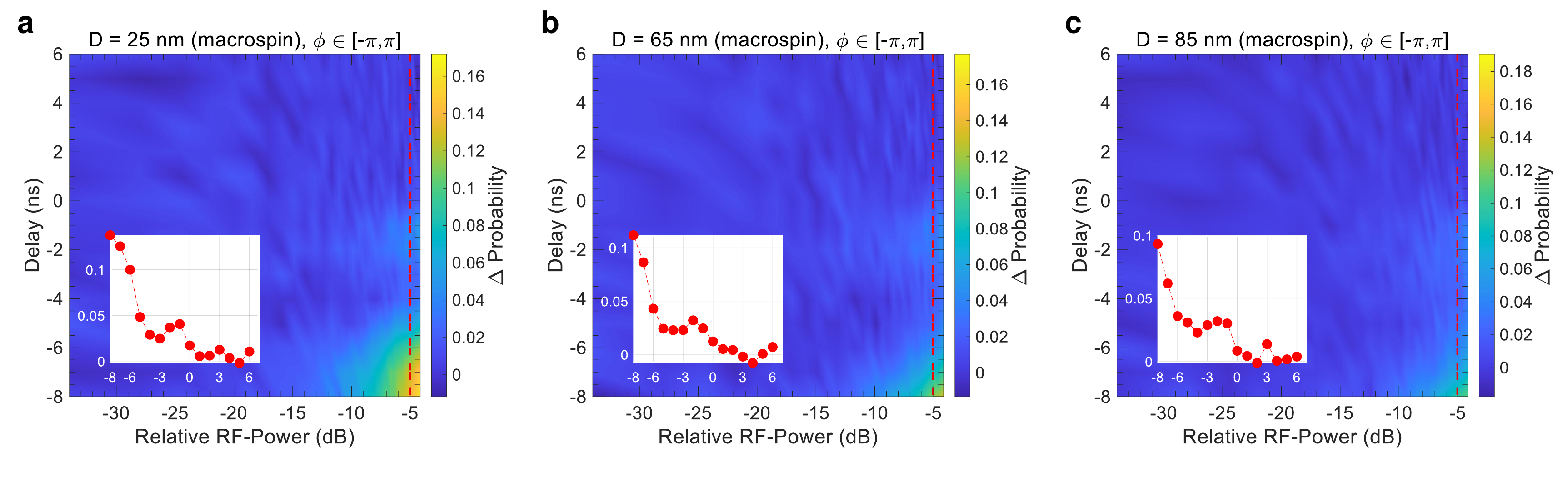}%
 \caption{\textbf{Change of switching probability as a function of of normalized RF power $\tilde{P}_\mathrm{RF}$ and delay $\tau$, with random phase $\phi\in[-\pi,\pi]$.} \textbf{a} $D=25$\,nm. \textbf{b} $D=65$\,nm. \textbf{c} $D=85$\,nm.
\label{fig:prob_diam}}%
 \end{figure*}
 
\subsection{Switching probability: additional figures}
In \cref{fig:prob_diam}, the change in switching probability is shown as a function of normalized RF-power $\tilde{P}_\mathrm{RF}$ -- defined relative to the minimal theoretical power required to switch the magnetic state with a DC current in the absence of thermal fluctuations (T = 0 K) -- and the delay $\tau$ of the DC pulse with respect to the RF pulse for $\phi\in[-\pi,\pi]$ and multiple device diameters. The diagrams look qualitatively similar; in particular, lower delay $\tau$ and higher relative power show higher enhancement of switching probability.

\bibliography{refer_sup}